\documentclass[reprint, amsmath, amssymb, aps, prd, nofootinbib]{revtex4-1}

\usepackage{graphicx}

\usepackage{bm}

\usepackage{hyperref}

\usepackage{physics}

\usepackage[caption=false]{subfig}

\usepackage{orcidlink}

\usepackage{colortbl}
\usepackage{hhline}
\definecolor{lightgray}{gray}{0.8}
\makeatletter
    \def\CT@@do@color{%
      \global\let\CT@do@color\relax
            \@tempdima\wd\z@
            \advance\@tempdima\@tempdimb
            \advance\@tempdima\@tempdimc
    \advance\@tempdimb\tabcolsep
    \advance\@tempdimc\tabcolsep
    \advance\@tempdima2\tabcolsep
            \kern-\@tempdimb
            \leaders\vrule
                    \hskip\@tempdima\@plus  1fill
            \kern-\@tempdimc
            \hskip-\wd\z@ \@plus -1fill }
\makeatother
\def\apj{\rm{ApJ}}
\def\apjl{\rm{ApJ}}

\def\mnras{\rm{MNRAS}}
\def\prd{\rm{PRD}}
\def\prl{\rm{PRL}}

\newcommand{\bham}{Institute for Gravitational Wave Astronomy \& School of Physics and Astronomy, University of Birmingham, Edgbaston, Birmingham B15 2TT, UK}
\newcommand{\IoA}{Institute of Astronomy, University of Cambridge, Madingley Road, Cambridge, CB3 0HA, UK}
\newcommand{\KICC}{Kavli Institute for Cosmology, University of Cambridge, Madingley Road, Cambridge, CB3 0HA, UK}
\newcommand{\DAMTP}{Department of Applied Mathematics and Theoretical Physics, Centre for Mathematical Sciences, University of Cambridge, Wilberforce Road, Cambridge, CB3 0WA, UK}

\begin{document}

\title{A GPU-accelerated semi-coherent hierarchical search for stellar-mass binary inspiral signals in LISA}

\author{Diganta Bandopadhyay
\orcidlink{0000-0003-0975-5613}}
\email{diganta@star.sr.bham.ac.uk}
\affiliation{\bham}

\author{Christopher J.\ Moore
\orcidlink{0000-0002-2527-0213}}
\email{cjm96@cam.ac.uk}
\affiliation{\IoA}
\affiliation{\DAMTP}
\affiliation{\KICC}

\date{\today}

\begin{abstract}
    Searching for gravitational waves from stellar-mass binary black holes with LISA remains a challenging open problem. 
    Conventional template-bank approaches to the search are impossible due to the prohibitive number of templates that would be required. 
    This paper continues the development of a hierarchical semi-coherent stochastic search, extending it to a full end-to-end pipeline that is then applied to multiple mock LISA data streams which include simulated noise.
    Particle swarm optimization is used as a stochastic search algorithm, tracking multiple maxima of a semi-coherent search statistic defined over source parameter space.     
    The pipeline is accelerated by the use of graphical processing units (GPUs).
    No prior information from observations by ground-based detectors is used; this is necessary in order to provide advance warning of the merger. 
    We find that the pipeline is able to detect sources with signal-to-noise ratios as low as $\rho\sim 17$, we demonstrate that these searches can directly seed coarse parameter estimation in cases where the search trigger is loud. 
    An example of how the false-alarm probability can be estimated for this type of GW search is also included. 
\end{abstract}

\maketitle

\section{Introduction}\label{sec:introduction}

The Laser Interferometer Space Antenna (LISA) will observe gravitational waves (GWs) from numerous astrophysical sources in the frequency range $\sim [10^{-4},10^{-1}]\, \rm{Hz}$ \cite{Colpi:2024}. 
Among these, there are two source classes in particular which are quiet, long lived, and  broadband: stellar-mass binary black holes (SmBBHs) and extreme mass ratio inspirals (EMRIs).
SmBBHs \citep{Sesana:2016} are the precursors of the LIGO-Virgo-Kagra (LVK) binary black holes in the early inspiral phase of their evolution, 
while EMRIs \citep{Gair:2017} are the capture of stellar-mass objects by a supermassive black hole. 
These sources pose particular data analysis challenges; the large number of observable orbits ($\gtrsim 10^{5}$) gives rise to tight constraints on the source parameters that control the waveform phase, requiring an extremely computationally expensive search to detect them \cite{Moore:2019,Gair:2004}.

The analysis of GW data is conventionally split into two phases: search (i.e.\ detection) and parameter estimation (i.e.\ characterizing the source). 
The clear distinction between the two phases works well for the current LVK compact binary coalescence events as these signals are well separated in time. 
However, the LISA data-stream will contain many overlapping and long-lived signals from different source-types, and therefore inter-source correlations cannot generally be ignored. An additional complicating factor is that there will be no signal-free LISA data.
This calls for a `global' approach to simultaneously search for and characterize all of the sources present.  
Early prototype implementations of `global-fit' algorithms have been presented in Refs.~\cite{Katz:2024,Littenberg:2023}.
However, due to their complexity, long-lived SmBBH and EMRI signals have not as of yet been included within these prototype global-fits, either in the search or parameter estimation parts. 
Parameter estimation of SmBBH \cite{Toubiana:2020, Buscicchio:2021, Klein:2022, Digman:2023} and EMRI \cite{Speri:2023,Burke:2024} signals have been tackled only in isolation (i.e.\ one signal at a time) and with very restrictive priors on key parameters. 
The search for these sources remains a key open problem.

In this work we continue developing a semi-coherent approach to the search for SmBBH signals, building on the work in Ref.~\cite{Bandopadhyay:2023}. 
Instead of initially trying to match the entire LISA time series against a single coherent waveform model, semi-coherent methods split up both the model and data into small segments (either in time or, as in this case, frequency) attempting to match them locally. 
The approach adopted here is inspired by methods developed for searching for continuous GWs (e.g.\ from spinning asymmetric neutron stars) in LVK data \citep{Riles:2023,D’Onofrio:2023,Ashton:2018}. 
The underlying reason for the large computational costs of a coherent analysis is the same for the SmBBH searches as it is for the LVK continuous wave searches; the long signal duration leads to extreme sensitivity to parameters that control the GW phase. We adopt a hierarchical search scheme, gradually increasing the level of coherence (achieved by decreasing the number of segments) to narrow in on the true parameters of the source; as the number of segments decreases so does the false-alarm probability.As the number of segments is increased (and the segments get correspondingly shorter in time/frequency) there is a greater chance for a model to be fit to noise in the data, which decreases the sensitivity of the search and increases the threshold SNR. Although our search pipeline does produce a crude posterior on the source parameters as an output, the focus here is very much on the search phase and identifying the presence of a source and estimating its statistical significance. We envision the crude posterior produced being used to seed a subsequent high-quality parameter inference with narrow priors. 

Currently, the only mature search strategy for SmBBH signals is an archival search. 
This approach involves waiting years to decades until after the source has coalesced and (hopefully, assuming one is operating at the time) been observed by a ground-based GW detector.
Given the exquisite parameter estimation results expected from the next generation of ground-based instruments, it will then be possible to go back and search in the archival LISA data with greatly restricted priors and an associated reduction in computational cost \cite{Wong:2018,Ewing:2021,Wang:2024}. 
However, Ref.~\cite{Wang:2024} has shown that the computational costs are still large when including the effects of eccentricity.
Although useful, archival searches are not able to provide advanced warning of the SmBBH merger for either ground-based GW or electromagnetic observatories looking for signatures of gas rich mergers \cite{deMink:2017,Yi:2019}.
Therefore, there is also a need to develop a `blind search' (i.e.\ one which uses just LISA data) which is the goal of this work.
Such a SmBBH search is a critical requirement for pre-merger alerts to ground-based observatories (science objective 3.4.3, Ref.~\cite{Colpi:2024}). 
Additionally, the methods developed here for SmBBHs may also be applied to EMRIs where an archival search is impossible.
It should be noted that the blind search is expected to be less sensitive than the archival search because of the much larger parameter space to be covered.

Besides Ref.~\cite{Bandopadhyay:2023}, another approach to a hierarchical semi-coherent search has been developed by Ref.~\cite{Fu:2024}.
This study used a slightly different search statistic which analytically maximises over a number of extrinsic parameters, reducing the dimensionality and thus volume of parameter space that must be searched. 
They also used a smaller number of semi-coherent segments resulting in a lower detection threshold for the signal-to-noise ratio (SNR) but at an increased computational cost compared to the method presented here.
Also, for computational reasons, Ref.~\cite{Fu:2024} restricted the analysis to a single TDI channel, whereas this work uses all three.
Additionally, other authors have been exploring machine learning based approaches for EMRI or SmBBH detection and search \cite{Zhang:2024,Zhang:2022}.
Although these show some promise in detecting the presence of a signal, currently there are difficulties in extracting estimates for the source parameters and the threshold SNR is currently much higher than with semi-coherent methods. 

Search strategies for some of the other astrophysical sources that we expect to observe in LISA are much more mature. 
The late inspiral and merger of supermassive black hole binaries (MBBH) are loud GW sources that are one of the main targets for the LISA mission, these signals enter the frequency bucket on the lower end and merge at $\sim 10^{-3}\, \mathrm{Hz}$ within a timescale of days to months. 
GWs from these sources are broadband, but localized in time. Search and parameter estimation for MBBH merger signals in LISA data has been demonstrated in \citep{Katz:2024,Weaving:2024,Cornish:2007}, with work ongoing to evaluate the pre-merger detectability of MBBH signals \citep{Ruan:2024,Dal_Canton:2019}. 
Numerous double white-dwarf (DWD) sources will also be observed by LISA, these quasimonochromatic sources will show up as very narrow features in the frequency spectrum, overlapping with MBBH mergers. 
These continuous-wave like sources are long lived but are localized in a narrow range of frequencies. 
Searches for individual DWD signals have been demonstrated in several studies, including Refs.~\citep{Strub:2023,Littenberg:2011}, with recent works demonstrating the joint search and characterization of MBBH and DWD sources \cite{Katz:2024,Strub:2024}. 

In our previous work we demonstrated the feasibility of a semi-coherent search on simulated SmBBH signals with zero noise \cite{Bandopadhyay:2023}. 
In this paper, the semi-coherent search is further developed in the presence of simulated noise. Particle swarm optimization (PSO) is used as the stochastic search algorithm over source parameter space, however it is here improved to allow for tracking the multiple maxima that are expected when noise is present in the data stream 
(this feature will also be vital when such methods are applied to the EMRI search; see, for example, Ref.~\cite{Chua:2022}). We also show that in the final stage of the search PSO particles can smoothly transition into Markov-chain Monte-Carlo (MCMC) walkers for parameter estimation.
Searching over noisy data is computationally expensive, and this is solved here by hardware acceleration using graphical processing units (GPU). 

This paper is structured as follows.
Sec.~\ref{sec:search_statistics} defines the semi-coherent search statistics and discusses their properties in the presence of noise. 
Sec.~\ref{sec:search} summarizes the structure of the search and details of PSO, including the changes made to adapt it to multi-peak optimization. 
Sec.~\ref{sec:Details} details the analysis pipeline, including the waveforms used, the LISA response, hardware acceleration etc. 
In Sec.~\ref{sec:search_statistic_statistical_properties} the sampling distributions of the semi-coherent search statistic discussed in Sec.~\ref{sec:search_statistics} are verified, additionally a probability-probability plot is computed to check the statistical properties of the semi-coherent likelihood. The results of several successful mock searches are presented in Sec.~\ref{sec:search_results}.
Importantly, Sec.~\ref{sec:search_results} also includes a discussion of how the false alarm probability (FAP) of a stochastic search can be estimated.
Sec.~\ref{sec:computational_cost} discusses the computational costs and hardware used to carry out the analyses in this paper.
Sec.~\ref{sec:discussion} contains concluding remarks.

\section{Search Statistics}\label{sec:search_statistics}

This section briefly summarizes and presents the key expressions for the semi-coherent data analysis quantities used in this study.
First, the coherent versions of these quantities are introduced in Sec.~\ref{subsec:coherent_stats} before their semi-coherent generalizations are introduced in Sec.~\ref{subsec:semicoherent_stats}.
This section builds on what was introduced in Ref.~\cite{Bandopadhyay:2023} and more details and further motivation for these definitions can be found there.

\subsection{Coherent search statistic} \label{subsec:coherent_stats}

Throughout this paper, the analysis is conducted in the frequency domain. 
All time series (both models and data) are assumed to be represented in the frequency domain, unless explicitly stated otherwise. The dependence on frequency is suppressed in order to avoid clutter in our notation. 

The main reason for doing GW data analysis in the frequency domain is that the usual noise-weighted inner product between signals $a$ and $b$ takes the form of a simple integral, 
\begin{equation} \label{eqn:inner_prod}
    \bra{a}\ket{b} = \sum_{\alpha} 4\mathrm{Re}\bigg\{\int_{f_{\rm min}}^{f_{\rm max}} \mathrm{d}f\; \frac{a_{\alpha}(f)b_{\alpha}^\dagger(f)}{S_{\alpha}(f)}\bigg\} ,
\end{equation}
in the case of stationary, Gaussian noise with one-sided noise power spectral density (PSD) $S_{\alpha}(f)$ in channel $\alpha$, where a dagger denotes complex conjugation. In the case of current ground-based detector networks, the index $\alpha$ labels different interferometers, in the case of LISA it labels the different noise-orthogonal time-delay interferometry (TDI) channels: $A$, $E$ and $T$ \cite{Armstrong:1999,Tinto:2021}.

The overlap is defined as the phase-maximized inner product,
\begin{equation} \label{eqn:overlap}
    \mathcal{O}(d, h) = \underset{\phi}{\text{max}} \bra{d}\ket{h(\theta) e^{i\phi}} ,
\end{equation}
where the exponential factor is phase shifting the single-mode (22-mode) waveform model $h(\theta)$, 
where $\theta$ denotes the vector of source parameters.
The optimal SNR $\rho$ for a model waveform $h(\theta)$ is defined as
\begin{equation}\label{eqn:SNR_coherent}
    \rho(\theta) = \sqrt{\bra{h(\theta)}\ket{h(\theta)}} ,
\end{equation}
The log-likelihood used for parameter estimation is defined as
\begin{align}\label{eqn:LogL_coherent}
    \log L(d|\theta) &= -\frac{1}{2}\bra{d-h(\theta)}\ket{d-h(\theta)} + \mathrm{const}, \\
    &= -\frac{1}{2}\bra{d}\ket{d} -\frac{1}{2}\bra{h}\ket{h} +\bra{d}\ket{h} , 
\end{align}
where $d$ denotes the measured data. 
Note, the normalization constant for the likelihood is discarded as it is only relevant when calculating the Bayesian evidence or when varying the noise PSD, neither of which are done in this study.
Generally, LISA data will contain a superposition of noise and many signals; i.e.\ $d = \sum_i h(\theta_{* i}) + n$, where $\theta_{* i}$ denote the true parameters of the $i^{\rm th}$ signal. 
However, here it is assumed that the data contains just a single source within the current region of parameter space being searched. 
This is likely to be a good assumption for SmBBH where the expected number of LISA sources is low \cite{Colpi:2024}. 
Furthermore, for broadband, long-duration signals like SmBBHs that contain a large number of orbits, the overlap between different signals is usually extremely small and can be safely neglected.

The aim of a GW search is to detect a source, provide a rough estimate of the parameters, and estimate its statistical significance, e.g.\ by calculating its false alarm probability (FAP). 
This information can then be used by a subsequent parameter estimation phase of the pipeline which refines the inference of $\theta_*$ by sampling from the full Bayesian posterior distribution $P(\theta|d) \propto L(d|\theta) \pi(\theta)$. 
In the search phase, the aim is not to perform sampling but instead to maximize a given detection statistic. 
Most detection statistics are based on the matched filter SNR which is defined as
\begin{equation} \label{eqn:SNR_mf}
    \tilde{\rho}(d,\theta) = \frac{\bra{d}\ket{h(\theta)}}{\sqrt{\bra{h(\theta)}\ket{h(\theta)}}} .
\end{equation}
Maximizing the matched filter SNR is equivalent to finding the maximum likelihood estimate for $\theta_*$. 
All GW searches maximize a function such as this, which is called the objective function.
Searches performed using data from the current ground-based detectors use variants of the matched filter, which efficiently maximize over parameters such as the time of merger and phase, reducing dimensionality of the parameter space (and thus the volume) for the search \cite{Usman:2016}. In practice, searches also use other information such as the quality of the fit using methods such as the $\chi^2$ statistic test \cite{Allen:2004}.

All of the quantities introduced in this section (the inner product in Eq.~\ref{eqn:inner_prod}, the overlap in Eq.~\ref{eqn:overlap}, the optimal SNR in Eq.~\ref{eqn:SNR_coherent}, the likelihood in Eq.~\ref{eqn:LogL_coherent}, and the matched filter SNR in Eq.~\ref{eqn:SNR_mf}) are referred to as \emph{coherent} quantities.
This is because they are defined in terms of a single integral of the smooth waveform model $h(\theta)$ over all values of frequency.
In the next section semi-coherent versions of these quantities are introduced.

\subsection{Semi-coherent search statistic} \label{subsec:semicoherent_stats}

While coherent search statistics have been successful in detecting compact binary coalescences in ground-based detectors, these sources have relatively small number of observable orbits (at most a few thousand, but often much less). 
Extending these search methods to sources with a much larger number of orbits (e.g.\ $10^5$--$10^6$) is a major challenge. 
In a previous study, we demonstrated an early proof-of-concept for a search using a likelihood function which used a semi-coherent matched filter \cite{Bandopadhyay:2023}, in this following section we recall the key expressions from our previous study and analytically study how the objective function of the search should change in the presence of noise. 

The essential feature of our semi-coherent method is that the inner product frequency integral is split into $N$ segments with boundaries in frequency $f_0<f_1<\ldots<f_N$ and where $f_0=f_{\rm min}$ and $f_N=f_{\rm max}$.
The inner product in the $n^{\rm th}$ segment is denoted using square brackets and is defined as
\begin{eqnarray}
    [a|b]^{N}_n = \sum_{\alpha} 4 \mathrm{Re} \left\{ \int_{f_{n}}^{f_{n+1}} \mathrm{d}f\; \frac{a_{\alpha}(f)b^\dagger_{\alpha}(f)}{S_{\alpha}(f)} \right\}.
\end{eqnarray}
The superscript capital $N$ denotes the (fixed) number of segments used while the (variable) subscript index $n=0,1,\ldots,N-1$ labels a specific segment.
These segmented inner products can be used to define a semi-coherent generalization of the overlap between two signals. 
Instead of just maximizing over the overall waveform phase (as was done in Eq.~\ref{eqn:overlap}), the semi-coherent overlap maximizes over the phase in each segment independently; no attempt is made to enforce any type of smoothness or phase continuity across the segment boundaries.
The semi-coherent overlap is defined as
\begin{eqnarray}\label{eq:semi-coherent_O}
    \hat{\mathcal{O}}_N(d,h) = \sum_{n=0}^{N-1} \underset{\phi_n}{\text{max}}\,[d|h(\theta)e^{i\phi_n}]^N_n .
\end{eqnarray}
For an intuitive representation of the semi-coherent phase maximized overlap, see Fig. 1 of Ref.~\cite{Bandopadhyay:2023}. By analogy with Eq.~\ref{eqn:LogL_coherent}, one can also define a semi-coherent version of the log-likelihood function using the segmented inner products;
\begin{equation}
    \log \hat{L}_N(d,\theta) = -\frac{1}{2}\bra{d}\ket{d}-\frac{1}{2}\bra{h}\ket{h} + \hat{\mathcal{O}}_N\big(d,h(\theta)\big) .
\end{equation}
This quantity is maximized with respect to the phase $\phi_n$ in each segment, this is simpler than attempting to marginalize out the phase. 
Therefore, the semi-coherent log-likelihood $\log \hat{L}_N(d,\theta)$ is a function of all the source parameters $\theta$ except the phase.

Our search will maximise a semi-coherent matched-filter statistic. In this paper we use the semi-coherent search statistic proposed in Ref.~\cite{Chua:2017}, which is the natural semi-coherent generalization of the coherent matched filter in  Eq.~\ref{eqn:SNR_mf}:
\begin{align}
    \Upsilon_N(d,\theta) &= \sum_{n=0}^{N-1} \frac{x_n^2}{[h(\theta)|h(\theta)]^{N}_n} ,\\
    \mathrm{where} \quad x_n &= \underset{\phi_n}{\text{max}}\,[d|h(\theta)e^{i\phi_n}]^N_n . \nonumber
\end{align}
The functional form of this definition can be understood by recognizing that each term in the summation corresponds to a phase-maximized $\tilde{\rho}^2$ in that segment.

In our notation semi-coherent quantities are denoted either with a hat (e.g.\ $\hat{L}$) and/or with a subscript capital $N$ (e.g.\ $\Upsilon_N$) which also indicates the number of segments used. 
The number of segments measures the adjustable level of coherence used in the analysis. If $N=1$ then we recover a fully coherent quantity. For example, $\Upsilon_{N=1}$ is equal to the square of the phase-maximized value of the coherent detection statistic $\tilde{\rho}^2$.

There is some freedom in how the segment frequency boundaries $f_n$ (for $n=1,2,\ldots,N-1$) are chosen.
As discussed in Sec.~2 of Ref.~\cite{Bandopadhyay:2023}, possible choices include spacing these uniformly (or log-uniformly) in frequency. However, here the segment boundaries are chosen such that each of the $N$ segments contains an equal amount of squared SNR. 
As a consequence of this choice, the segment boundaries are themselves functions of the source parameters $\theta$; changing the parameters of a source effects not only the total SNR, but also how SNR accumulates over the frequency band.
Therefore, $f_n$ is reevaluated for every new value of $\theta$ every time a semi-coherent quantity is to be evaluated. This differs from our previous work where the segment boundaries were fixed \cite{Bandopadhyay:2023}.
For a linearly chirping quasi-monochromatic source (i.e.\ $\dot{f}_{\rm GW}\approx\mathrm{const} \ll f_{\rm{GW}}/T$), and in the case of a white instrumental noise PSD (i.e.\ $S_{\alpha}(f)\approx\mathrm{const}$), choosing the segment boundaries based on an equal squared SNR per segment can be shown to be equivalent to splitting the signal in $N$ equal length segments in the time domain (see Appendix \ref{app:linear_chirp}).

\subsection{Sampling distributions}\label{sec:Search_statistics_variation_under_noise}

\begin{figure*}[t]
    \centering
    \includegraphics[width=\textwidth]{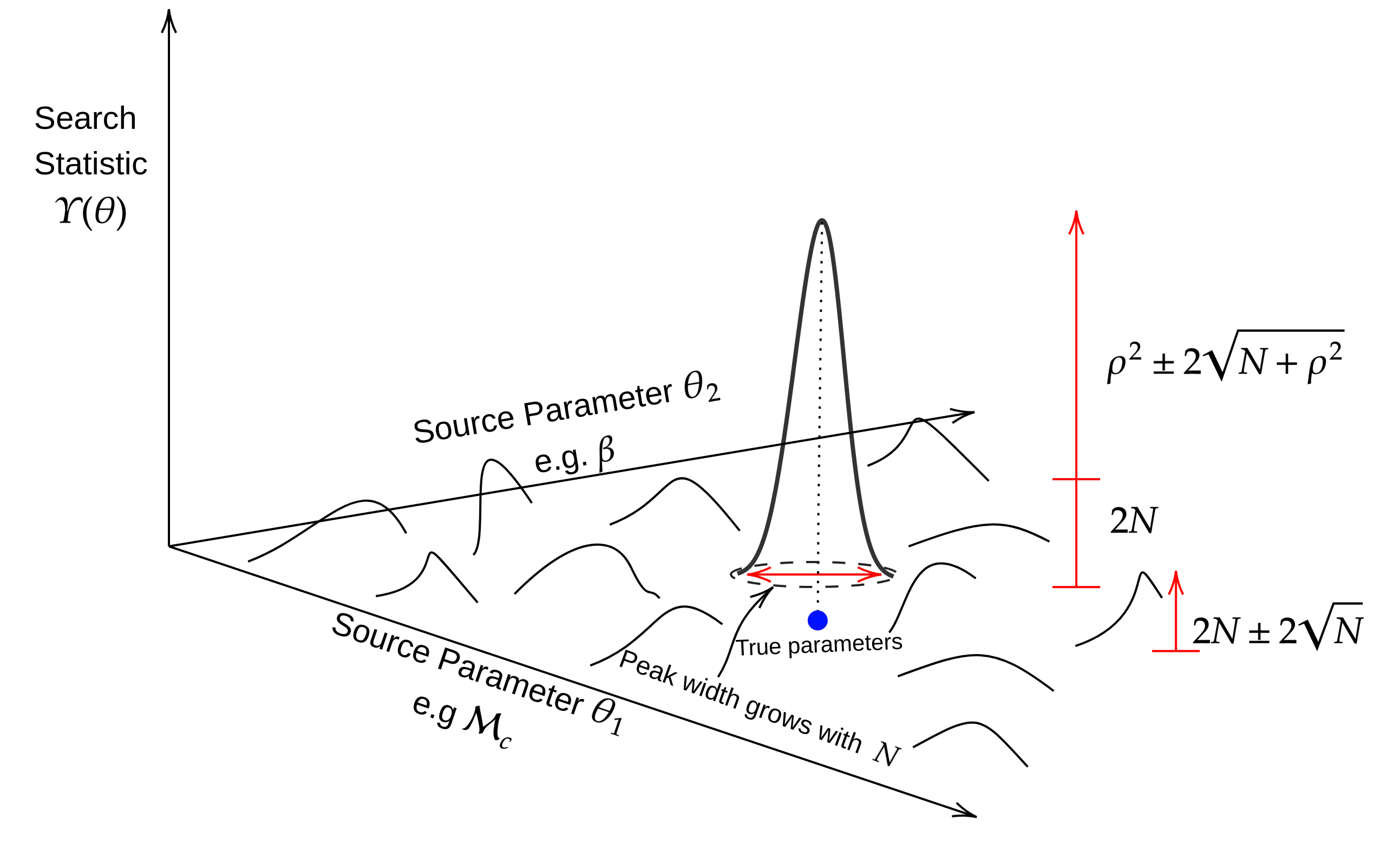}
    \caption{Sketch illustrating the variation of the semi-coherent search statistic $\Upsilon_N(d,\theta)$ with $N$ segments over the parameter space $\theta$. 
    The statistic peaks near the true source parameters, $\theta_*$, where the expected value $(2N+\rho^2)$ and standard deviation $(2\sqrt{N+\rho^2})$ of $\Upsilon_N(d,\theta_*)$ depends on the optimal squared SNR of the source and the number of semi-coherent segments, $N$. 
    These values for the expectation and variance, given in Eq.~\ref{eqn:CLTdist_signal}, are illustrated by the long red arrow. 
    Away from the true parameter values, the search statistic fluctuates due to chance partial overlaps between the random noise and template. 
    The values of the expectation $(2N)$ and standard deviation $(2\sqrt{N})$ of the statistic away from the true source parameters, given in Eq.~\ref{eqn:CLTdist_noise}, are illustrated by the short red arrow. }
    \label{fig:Search_statistic_schematic}
\end{figure*}
The search statistics (both coherent $\tilde{\rho}(d,\theta)$ and semi-coherent $\Upsilon_N(d,\theta)$) are functions of the noisy data $d$ and are therefore random variables. The random element arises from the inclusion of noise in $d$, due to this the value of $\Upsilon_N(d,\theta)$ or $\tilde{\rho}(d,\theta)$ will vary at a fixed $\theta$ if the noise realization in the data ($d$) changes. If they are to be used in a search, it is important that the sampling distributions of these statistics (i.e.\ their probability distributions) are properly understood as these underpin the false alarm and detection probabilities for the search. 
The sampling distributions need to be known both in the case when the data contains only noise, i.e.\ when $d=n$, and also when the data contains the signal being searched for, $d=n+h(\theta)$.

It is well known that the standard matched filter $\tilde{\rho}(d,h)$ is normally distributed with unit variance and a mean given by the optimum SNR $\rho$ (if there is no signal present, then the mean is zero).
If the coherent detection statistic is maximized over phase (i.e., $\phi_{\rm max}(\tilde{\rho}
)$) then it can be shown that this statistic follows either a Rayleigh (in the only-noise case) or a Rice distribution (signal case) \cite{Buonanno:2003}. Since $\Upsilon_1 $ is equivalent to $\phi_{\rm max}(\tilde{\rho})^2$, it follows a standard chi-squared distribution with 2 degrees of freedom $\chi_2^2$ (in the only-noise case) or a non-central chi-squared distribution with 2 degrees of freedom and a non-centrality parameter $\lambda=\rho^2$,  $\chi_2^2(\lambda)$ (in the signal case), both distributions have unit scale parameter.  

It is not possible to derive the corresponding analytic distributions for the semi-coherent search statistic $\Upsilon_N$ for finite $N$.
However, in the limit of many segments, the following approximations to these distributions were derived in Ref.~\cite{Chua:2017} using the central limit theorem (CLT). 
In the presence of noise only, it was shown that approximately 
\begin{align}\label{eqn:CLTdist_noise}
    \Upsilon_N &\sim \mathcal{N}(\mu_0,\sigma_0^2) , \\
    \mathrm{where}\; \mu_0&=N\mu_k\; \mathrm{and}\;\sigma^2_0=2N\sigma^2_k. \nonumber
\end{align}
And in the presence of noise and signal, it was shown that approximately 
\begin{align}\label{eqn:CLTdist_signal}
    \Upsilon_N &\sim \mathcal{N}(\mu_1,\sigma_1^2) , \\
    \mathrm{where}\; \mu_1&=A^2+N\mu_k\; \mathrm{and}\;\sigma^2_1=2N\sigma^2_k+4A^2. \nonumber 
\end{align}
Where $(\mu_k,\sigma_k) = (2.00,\sqrt{2})$ and $A = \rho$. These are referring to the expected distributions of $\Upsilon_N$ when it is evaluated with many independent realizations of the noise in the data-stream. Because these are based on the CLT, they are expected to be a good approximation in the limit of large $N$.
In Ref.~\cite{Chua:2017}, these approximate distributions were tested by drawing and combining random samples from the analytic Rayleigh and Rice distributions on which they based and Eqs.~\ref{eqn:CLTdist_noise} and \ref{eqn:CLTdist_signal} were found to provide good approximations for $N\gtrsim 70$. 

In this paper, these approximate distributions are tested in a more realistic way by simulating mock LISA time series containing Gaussian noise with and without signals. The numerical results are described in Sec.~\ref{subsec:results_upsilon_dist}, the approximate distributions derived from the CLT are shown to be a good approximation even for $N$ as a low as 10. 

These probability distributions allow us to calculate the FAP of the search statistic at a fixed point in parameter space. However this is not indicative of the FAP for the whole search as that covers a large region in $\theta$. 
The FAP is calculated by simulating searches over a large number of noise and noise + waveform data streams  \citep{Usman:2016,Nitz:2017}. A similar method is used to make a limited estimate of the FAP for this search in section \ref{sec:search_results}.

At a higher number of segments, lower density of templates covering $\theta$ is traded for a higher FAP. Practically the higher FAP translates to the `noise' maximas in $\Upsilon_N$ being of a greater height and more comparable to the `true' peak; this is described by equations \ref{eqn:CLTdist_noise} and \ref{eqn:CLTdist_signal} which is illustrated by Fig.~\ref{fig:Search_statistic_schematic}. Since the height of the `true' peak is proportional to the SNR of the injected signal, the noise peaks can be problematic for signals at low SNR. This is relevant for the SmBBH inspirals as these sources are expected to be quiet \citep{Ewing:2021,Moore:2019}. This limits the maximum number of segments we can use for the search; using a very high number of segments increases the height of the noise peaks, thereby increasing the chance that a search may mistake such peaks in $\Upsilon_N$ as a signal. 

\section{Search Structure}\label{sec:search}

The aim of the search is to maximise the matched-filter statistic with respect to $\theta$. 
In LVK searches for compact binary coalescence signals a template bank approach is used. 
This involves generating a set of waveforms $\{h_0,h_1,\hdots,h_{N-1}\}$ at predetermined locations in parameter space $\{\theta_0,\theta_1,\hdots,\theta_{N-1}\}$, and computing the (coherent) inner product between the data and every template in the bank (maximized over phase and time offsets).
The search identifies the template with the largest value of the statistic.
As discussed in the introduction, for long-lived signals such as LISA SmBBH inspirals this procedure is not possible due to the number of templates that would be required (see, e.g., Ref.~\cite{Moore:2019}).

Our proposed solution to this problem involves three changes with respect to the LVK searches. The first is the switch from the coherent to the semi-coherent inner product described in Sec.~\ref{subsec:semicoherent_stats}. The second is the hierarchical nature of the search where the coherence is gradually increased. Finally, the third is the use of a stochastic optimization method to maximize the detection statistic, instead of the deterministic template bank; this is described in this section.

This section describes the broad conceptual features of the search. Technical details specific to this particular implementation are presented in Sec.~\ref{sec:Details}.

\subsection{Particle swarm optimization}\label{subsec:PSO}

In this study, PSO is used as the stochastic optimization algorithm \cite{Shi:1998}. 
PSO uses a swarm of particles (similar to the walkers from ensemble MCMC methods) to explore the parameter space $\theta$ and maximize a given objective function. 
In our implementation, 60000 particles are used.
In this case the particle positions are initialized by drawing from the prior and the objective function is taken to be the semi-coherent search statistic $\Upsilon_N$.

PSO is an iterative algorithm in which the positions of each particle are updated at each iteration according to the rule
\begin{equation}
    \theta_{p,i}^{\mu} = \theta_{p,i-1}^{\mu} + v_{p,i}^{\mu} .
\end{equation}
Here $i$ is the iteration number, $p$ is the particle number and $\mu$ is the component of the parameter vector being updated. The particle velocities are calculated using Eq. \ref{eqn:PSO_velocity_rule} (with an additional intermediate step in Eq. \ref{eqn:PSO_min_velocity}),

\begin{eqnarray}\label{eqn:PSO_velocity_rule}
   u^{\mu}_{p,i} = \Omega v_{p,i-1}^{\mu} +  \Phi_{P}(r_P)^{\mu}_{p,i} (\Psi_{p,i-1}^\mu-\theta_{p,i-1}^{\mu}) +\\ 
   \Phi_{G}(r_G)^{\mu}_{p,i}(\Xi_{i-1}^\mu-\theta_{p,i-1}^{\mu}) . \nonumber
\end{eqnarray}

The parameters $\Omega$, $\Phi_P$ and $\Phi_G$ control the gross behaviour of the swarm.
$\Omega$ controls the inertial component of a particles velocity.
The coefficients $\Phi_P$ ($\Phi_G$) multiply vectors that point from a particles current position to the best position that particle (the swarm) has previously visited. 
$\Psi_{p,i}^\mu$ denotes the best position the particle $p$ has visited up to iteration $i$, whereas $\Xi_i^\mu$ is best position visited by the swarm up to iteration $i$. This follows the notation in Ref.~\cite{Bandopadhyay:2023}.
The velocity update rule is stochastic because the $(r_P)^{\mu}_{p,i}$ and $(r_G)^{\mu}_{p,i}$ are uniformly distributed random numbers in the interval $(0,1)$ that are drawn independently for each particle, for each dimension and for each iteration. 
As discussed in more detail in Ref.~\cite{Bandopadhyay:2023}, the swarm can be made to behave in an exploratory manner (i.e.\ better at looking for new peaks) by decreasing $\Phi_G$ and increasing $\Phi_P$ and $\Omega$ or in an exploitative manner (i.e.\ better at climbing existing peaks) with the opposite choices. For a review, see Ref.~\cite{Nickabadi:2011}.

It was also found to be helpful to enforce a minimum velocity in each parameter, this prevents premature convergence to local maxima. This is referred to as velocity clamping. At each iteration the velocity calculated in Eq.~\ref{eqn:PSO_velocity_rule} is modified according to
\begin{equation}\label{eqn:PSO_min_velocity}
        v_{p,i}^{\mu} = \mathrm{max}(\epsilon^{\mu}, |u^{\mu}_{p,i}|) \frac{u^{\mu}_{p,i}}{|u^{\mu}_{p,i}|}.
\end{equation}
This enforces a minimum velocity in each parameter at each iteration via the clamping parameters $\epsilon^\mu$. 
The values of the hyper-parameters are determined empirically and are changed throughout the search; the values used here are given in Appendix \ref{app:PSO_params}. 

The search proceeds in a hierarchical manner, with the level of coherence demanded gradually increasing as the swarm converges. 
This is achieved by decreasing in stages the number of segments $N$ used in the objective function $\Upsilon_N$. 
At each hierarchical stage, $\Upsilon_N$ is optimized until the swarm has reached a convergence criteria (see Appendix \ref{app:PSO_params}) at which point the objective function is changed to $\Upsilon_{N'}$ (where $N'<N$). This sequence of decreasing $N$ values is referred to as a ladder. Optimization at each level is repeated until the final step in the ladder ($N=1$) has been reached and optimized. The coherent search statistic ($\Upsilon_{N=1}$) has the lowest FAP out of all levels in the ladder.
\subsection{Multi-Peak tracking}

As discussed in Sec.~\ref{sec:Search_statistics_variation_under_noise}, using a high number of segments increases the FAP of a search using the statistic $\Upsilon_N$. 
However, a large $N$ is needed to efficiently find and localise signal peaks early in the search, especially when using broad priors. 

The hierarchical search method presented here aims to solve this problem by starting with a large number of segments (initially accepting the associated high FAP) and tracking multiple source candidates through a sequence of decreasing values of $N$.
As false positive candidates identified early in this process are tracked, the peak in the $\Upsilon_N$ surface becomes less significant (i.e. consistent with the background noise distribution) as $N$ is decreased; this allows the false positive candidate to be vetoed later in the process. 
Although the hierarchical search has a high FAP during the early stages, spurious sources are eliminated by the time $\Upsilon_{N=1}$ has been reached. 
Since $\Upsilon_{N=1}$ has similar FAP to $\tilde{\rho}$ for long signals, any signal that has a low FAP in $\Upsilon_{N=1}$ can be treated as being a statistically significant source (at the same level as if it were a trigger from searching using $\tilde{\rho}$). 

To do this we need an optimizer that can track a large number of optima throughout the search, accounting for the possibility of changing (both increasing and decreasing) the number of optima being optimized. We do this using an implementation of a multi-swarm PSO algorithm. 
Similar ideas have been explored for multi-objective optimization in \cite{Kennedy:2000}. 
The optimizer is initialized with an (arbitrary) number of swarms each containing the same number of particles. Each swarm is evolved using the usual PSO rule described in Sec.~\ref{subsec:PSO}. 
Once each swarm has converged on the $\Upsilon_{N_{\rm{high}}}$ surface, $k$-means clustering is run on the positions of all particles on all swarms to (re)group them. 
This allows for the merging of swarms which are optimizing the same peak or for the creation of new swarms when additional peaks are detected. 
The total number of particles is conserved throughout the search. 
The search statistic is then changed to $\Upsilon_{N'}$ ($N'<N$) as we move down a rung of the ladder.
The search and re-clustering processes repeat until all swarms have converged on the $N=1$ surface. 
The positions and $\Upsilon_{N=1}$ function values at the end of the search are the main results.

The use of multiple-swarms and $k$-means clustering is a significant development compared to our earlier work.
See Appendix \ref{app:PSO_params} for the values of all the multi-swarm PSO algorithm hyper-parameters used for the search. 

Note that PSO is not invariant with respect to a change of parameters $\theta\rightarrow\theta'(\theta)$.
In particular, PSO is not invariant with respect to rotations in the parameter space.
This can lead to undesirable features in the particle locations when optimizing simple, approximately Gaussian-looking peaks; see, for example, Fig.~2 of Ref.~\cite{Prasad:2012} and Ref.~\cite{Spears:2010} for a discussion.
We have found that this can be problematic for our search especially when  $\Upsilon_N$ has non-linear correlations in parameter space. 
At the moment, this has been overcome simply by increasing the total number of particles used.
However, we note there exist rotationally invariant versions of the PSO algorithm (e.g.\ \cite{Wilke:2007}), these may help alleviate these problems. However, these algorithms are more complicated and have additional tuning parameters which need to be optimized. We do not investigate such algorithms here.

\subsection{Transition to MCMC inference}\label{subsec:automated_PE}

The final aim of a search process is to provide rough estimates for the source parameters of the detected signals, these can then be used to `seed' parameter estimation. At this early stage of development, we do this by directly initializing MCMC particles using some particles from the final state of the search. After the final search stage, a small number of particles from each swarm (that has passed a detection threshold) are extracted and used as initial positions for MCMC walkers. The posterior distribution sampled at this final stage uses the phase-maximized coherent likelihood $\log\hat{L}_1$. 
We have verified that for long-lived signals this produces posteriors extremely similar to the standard coherent likelihood \cite{Bandopadhyay:2023}. 

Each PSO particle that transitions into an MCMC walker draws a luminosity distance $D_L$ from the prior, the search statistic analytically maximizes over this parameter so it must be drawn for  each particle (or walker) at the point of transition. 

We use \texttt{Zeus} to sample the posterior associated with $\log \hat{L}_1$.
\texttt{Zeus} is an MCMC sampler which uses ensemble slice sampling to efficiently traverse highly correlated parameter spaces \cite{Karamanis:2021,Karamanis:2020}. We found this to be particularly well-suited to some of the non-linear degeneracies and correlations present in the likelihood of SmBBH signals.

\section{Search Implementation Details}\label{sec:Details}

\begin{figure*}[ht!]

\tikzset{every picture/.style={line width=0.75pt}} 

\begin{tikzpicture}[x=0.75pt,y=0.75pt,yscale=-1,xscale=1]

\draw   (174.82,21.2) .. controls (174.82,16.78) and (178.4,13.2) .. (182.82,13.2) -- (272,13.2) .. controls (276.42,13.2) and (280,16.78) .. (280,21.2) -- (280,45.2) .. controls (280,49.62) and (276.42,53.2) .. (272,53.2) -- (182.82,53.2) .. controls (178.4,53.2) and (174.82,49.62) .. (174.82,45.2) -- cycle ;
\draw   (177.22,123.6) .. controls (177.22,119.18) and (180.8,115.6) .. (185.22,115.6) -- (274.4,115.6) .. controls (278.82,115.6) and (282.4,119.18) .. (282.4,123.6) -- (282.4,147.6) .. controls (282.4,152.02) and (278.82,155.6) .. (274.4,155.6) -- (185.22,155.6) .. controls (180.8,155.6) and (177.22,152.02) .. (177.22,147.6) -- cycle ;
\draw    (226.4,52.6) -- (226.4,113.87) ;
\draw [shift={(226.4,115.87)}, rotate = 270] [color={rgb, 255:red, 0; green, 0; blue, 0 }  ][line width=0.75]    (10.93,-3.29) .. controls (6.95,-1.4) and (3.31,-0.3) .. (0,0) .. controls (3.31,0.3) and (6.95,1.4) .. (10.93,3.29)   ;
\draw   (178.82,198) .. controls (178.82,193.58) and (182.4,190) .. (186.82,190) -- (276,190) .. controls (280.42,190) and (284,193.58) .. (284,198) -- (284,222) .. controls (284,226.42) and (280.42,230) .. (276,230) -- (186.82,230) .. controls (182.4,230) and (178.82,226.42) .. (178.82,222) -- cycle ;
\draw    (228,156.67) -- (228,187.47) ;
\draw [shift={(228,189.47)}, rotate = 270] [color={rgb, 255:red, 0; green, 0; blue, 0 }  ][line width=0.75]    (10.93,-3.29) .. controls (6.95,-1.4) and (3.31,-0.3) .. (0,0) .. controls (3.31,0.3) and (6.95,1.4) .. (10.93,3.29)   ;
\draw    (284,218.6) -- (459.28,218.6) ;
\draw   (408.62,165.8) .. controls (408.62,161.38) and (412.2,157.8) .. (416.62,157.8) -- (505.8,157.8) .. controls (510.22,157.8) and (513.8,161.38) .. (513.8,165.8) -- (513.8,189.8) .. controls (513.8,194.22) and (510.22,197.8) .. (505.8,197.8) -- (416.62,197.8) .. controls (412.2,197.8) and (408.62,194.22) .. (408.62,189.8) -- cycle ;
\draw    (459.2,219.2) -- (459.2,208.7) -- (459.2,200.07) ;
\draw [shift={(459.2,198.07)}, rotate = 90] [color={rgb, 255:red, 0; green, 0; blue, 0 }  ][line width=0.75]    (10.93,-3.29) .. controls (6.95,-1.4) and (3.31,-0.3) .. (0,0) .. controls (3.31,0.3) and (6.95,1.4) .. (10.93,3.29)   ;
\draw    (461.2,157) -- (461.2,125.65) ;
\draw    (321.33,127.4) -- (285.33,127.4) ;
\draw [shift={(283.33,127.4)}, rotate = 360] [color={rgb, 255:red, 0; green, 0; blue, 0 }  ][line width=0.75]    (10.93,-3.29) .. controls (6.95,-1.4) and (3.31,-0.3) .. (0,0) .. controls (3.31,0.3) and (6.95,1.4) .. (10.93,3.29)   ;
\draw    (227.2,229.47) -- (227.2,260.27) ;
\draw [shift={(227.2,262.27)}, rotate = 270] [color={rgb, 255:red, 0; green, 0; blue, 0 }  ][line width=0.75]    (10.93,-3.29) .. controls (6.95,-1.4) and (3.31,-0.3) .. (0,0) .. controls (3.31,0.3) and (6.95,1.4) .. (10.93,3.29)   ;
\draw   (178.02,271.6) .. controls (178.02,267.18) and (181.6,263.6) .. (186.02,263.6) -- (275.2,263.6) .. controls (279.62,263.6) and (283.2,267.18) .. (283.2,271.6) -- (283.2,295.6) .. controls (283.2,300.02) and (279.62,303.6) .. (275.2,303.6) -- (186.02,303.6) .. controls (181.6,303.6) and (178.02,300.02) .. (178.02,295.6) -- cycle ;
\draw    (231.2,303.6) -- (231.2,340.27) ;
\draw [shift={(231.2,342.27)}, rotate = 270] [color={rgb, 255:red, 0; green, 0; blue, 0 }  ][line width=0.75]    (10.93,-3.29) .. controls (6.95,-1.4) and (3.31,-0.3) .. (0,0) .. controls (3.31,0.3) and (6.95,1.4) .. (10.93,3.29)   ;
\draw   (174.82,355.73) .. controls (174.82,348.15) and (180.96,342) .. (188.55,342) -- (280.8,342) .. controls (288.39,342) and (294.53,348.15) .. (294.53,355.73) -- (294.53,396.92) .. controls (294.53,404.5) and (288.39,410.65) .. (280.8,410.65) -- (188.55,410.65) .. controls (180.96,410.65) and (174.82,404.5) .. (174.82,396.92) -- cycle ;
\draw    (229.4,411.2) -- (229.4,447.55) ;
\draw [shift={(229.4,449.55)}, rotate = 270] [color={rgb, 255:red, 0; green, 0; blue, 0 }  ][line width=0.75]    (10.93,-3.29) .. controls (6.95,-1.4) and (3.31,-0.3) .. (0,0) .. controls (3.31,0.3) and (6.95,1.4) .. (10.93,3.29)   ;
\draw   (177.02,457.49) .. controls (177.02,452.36) and (181.18,448.2) .. (186.31,448.2) -- (272.91,448.2) .. controls (278.04,448.2) and (282.2,452.36) .. (282.2,457.49) -- (282.2,485.36) .. controls (282.2,490.49) and (278.04,494.65) .. (272.91,494.65) -- (186.31,494.65) .. controls (181.18,494.65) and (177.02,490.49) .. (177.02,485.36) -- cycle ;
\draw   (321.62,114.8) .. controls (321.62,110.38) and (325.2,106.8) .. (329.62,106.8) -- (418.8,106.8) .. controls (423.22,106.8) and (426.8,110.38) .. (426.8,114.8) -- (426.8,138.8) .. controls (426.8,143.22) and (423.22,146.8) .. (418.8,146.8) -- (329.62,146.8) .. controls (325.2,146.8) and (321.62,143.22) .. (321.62,138.8) -- cycle ;
\draw    (427,125.6) -- (461.07,125.6) ;
\draw    (284.1,285.4) -- (333.28,285.4) ;
\draw [shift={(335.28,285.4)}, rotate = 180] [color={rgb, 255:red, 0; green, 0; blue, 0 }  ][line width=0.75]    (10.93,-3.29) .. controls (6.95,-1.4) and (3.31,-0.3) .. (0,0) .. controls (3.31,0.3) and (6.95,1.4) .. (10.93,3.29)   ;
\draw   (335.62,273.8) .. controls (335.62,269.38) and (339.2,265.8) .. (343.62,265.8) -- (432.8,265.8) .. controls (437.22,265.8) and (440.8,269.38) .. (440.8,273.8) -- (440.8,297.8) .. controls (440.8,302.22) and (437.22,305.8) .. (432.8,305.8) -- (343.62,305.8) .. controls (339.2,305.8) and (335.62,302.22) .. (335.62,297.8) -- cycle ;

\draw (192.02,18.8) node [anchor=north west][inner sep=0.75pt]   [align=left] {\begin{minipage}[lt]{49.86pt}\setlength\topsep{0pt}
\begin{center}
{\small Tile search }\\{\small prior}
\end{center}

\end{minipage}};
\draw (185.62,122.6) node [anchor=north west][inner sep=0.75pt]   [align=left] {\begin{minipage}[lt]{66.03pt}\setlength\topsep{0pt}
\begin{center}
PSO Optimize $\displaystyle \Upsilon _{N^{\mu }}$
\end{center}

\end{minipage}};
\draw (202.42,200.8) node [anchor=north west][inner sep=0.75pt]   [align=left] {\begin{minipage}[lt]{42.44pt}\setlength\topsep{0pt}
\begin{center}
$\displaystyle N\ =1?$
\end{center}

\end{minipage}};
\draw (426.42,166.4) node [anchor=north west][inner sep=0.75pt]   [align=left] {\begin{minipage}[lt]{49.58pt}\setlength\topsep{0pt}
\begin{center}
$\displaystyle \mu =\mu +1$
\end{center}

\end{minipage}};
\draw (230,72.6) node [anchor=north west][inner sep=0.75pt]    {$\mu =0$};
\draw (37,134.4) node [anchor=north west][inner sep=0.75pt]   [align=left] {{\footnotesize Gradually increasing }\\{\footnotesize level of coherence }\\{\footnotesize through the ladder}\\{\footnotesize  $\displaystyle N^{\mu } \ =\left\{N^{0} ,N^{1} ,\dotsc ,1\right\}$}\\{\footnotesize until reaching $\displaystyle N=\ 1$.}};
\draw (295.6,188.4) node [anchor=north west][inner sep=0.75pt]   [align=left] {No};
\draw (234,234) node [anchor=north west][inner sep=0.75pt]   [align=left] {Yes};
\draw (181.02,276) node [anchor=north west][inner sep=0.75pt]   [align=left] {\begin{minipage}[lt]{71.01pt}\setlength\topsep{0pt}
\begin{center}
{\small $\displaystyle \Upsilon _{1}  >$threshold?}
\end{center}

\end{minipage}};
\draw (30,275) node [anchor=north west][inner sep=0.75pt]   [align=left] {{\footnotesize Each candidate}\\{\footnotesize `source' (swarm) is subject }\\{\footnotesize to this noise threshold }\\{\footnotesize cutoff}};
\draw (190.2,355) node [anchor=north west][inner sep=0.75pt]   [align=left] {Seed MCMC \\with 100 best\\swarm particles};
\draw (201.8,460.2) node [anchor=north west][inner sep=0.75pt]   [align=left] { MCMC\\sampling};
\draw (336.62,108.6) node [anchor=north west][inner sep=0.75pt]   [align=left] {\begin{minipage}[lt]{52.05pt}\setlength\topsep{0pt}
\begin{center}
Re-cluster \\swarms
\end{center}

\end{minipage}};
\draw (238,308) node [anchor=north west][inner sep=0.75pt]   [align=left] {Yes};
\draw (292.6,260.4) node [anchor=north west][inner sep=0.75pt]   [align=left] {No};
\draw (362.62,277.6) node [anchor=north west][inner sep=0.75pt]   [align=left] {\begin{minipage}[lt]{37.31pt}\setlength\topsep{0pt}
\begin{center}
Discard
\end{center}

\end{minipage}};

\end{tikzpicture}

\caption{Schematic flowchart for the search pipeline within one $\mathcal{M}_c-t_c$ tile. The noise threshold used in this study for each swarm is $\Upsilon_1 \approx 100$, see top panel of Fig. \ref{fig:real_false_alarm_prob_combined_figure}.}
\label{fig:schematic_flowchart}
\end{figure*}
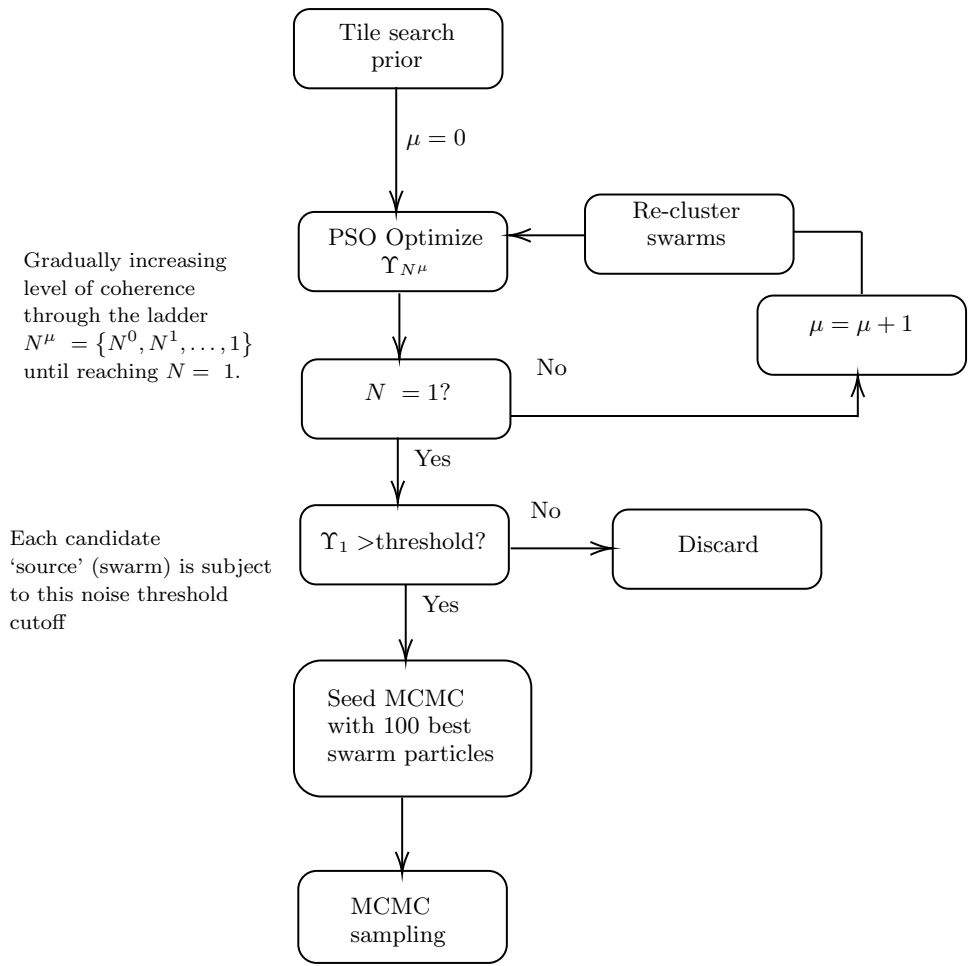

In this section the technical details of the search and analysis pipeline as implemented here are presented. This includes details about the waveform, instrument response and hardware acceleration used in this study. 

The \texttt{TaylorF2Ecc} waveform was used for injection, search and parameter estimation \cite{Moore:2016}. 
This is a frequency domain, post-Newtonian (PN) waveform approximant for the $(l,\abs{m})=(2,2)$ mode, which includes contributions from the orbital eccentricity to the GW phase. This PN waveform is expected to be sufficiently accurate for the search and recovery of the early inspiral signal \cite{Mangiagli:2019}. Binaries merging in the LIGO/VIRGO frequency band are expected to have radiated away much of the orbital eccentricity, however the early inspiral of these systems -- observable in LISA -- may retain eccentricity \cite{Fumagalli:2024}, which contributes significantly to the waveform phase. 
This \texttt{TaylorF2Ecc} waveform neglects the BH spins. The component spins of SmBBH systems will not be well measured by LISA, although the aligned spin contribution to the GW phase may be moderately significant \cite{Buscicchio:2021} and should be included in future work. 
This waveform has recently been widely used in the context of LISA; see, for example, Refs.~\cite{Garg:2024,Garg:2023}, similar waveforms are also used in Refs.~\cite{Wang:2024,Fu:2024}. 

Early parameter estimation studies for SmBBH signals in LISA have mostly been limited to the case of zero noise. 
This is partly because the inclusion of noise makes the integrand for the $\bra{d}\ket{h}$ term in the likelihood a non-smooth function of frequency, which renders common methods of numerically evaluating this integral, such as quadrature integration, invalid; for a detailed discussion see Appendix D of Ref.~\cite{Bandopadhyay:2023}. We are forced to evaluate the waveform and sum in Eq.~\ref{eqn:inner_prod} over the dense fast-Fourier transform (FFT) frequency grid. The number of points in this grid is $\sim [f_{\rm{max}}-f_{\rm{min}}]T_{\rm{obs}}$. Since SmBBH sources are broadband ($f_{\rm{max}}-f_{\rm{min}} \sim 10^{-1} \, \rm{Hz}$) and long lived ($T_{\rm{obs}} \sim $ years), this is typically millions of points. 
For other source types, such as MBBH (short duration, with $T_{\rm{obs}} \sim $ weeks) and DWD (narrow band, with $f_{\rm{max}}-f_{\rm{min}} \sim 10^{-6} \,  \rm{Hz}$), the number of frequency points is much smaller. 

A common method of accelerating the waveform evaluations is through the use of interpolation, evaluating the waveform on a sparse frequency grid and then interpolating onto the denser frequency grid where inner product or likelihood is computed. Similar approaches have previously been used in, for example, Refs.~\citep{Toubiana:2020,Katz:2020,Katz:2022}. A decomposition of the waveform into amplitude and phase, in the form $h(f)=A(f)e^{i \phi(f)}$ is convenient for interpolation. The amplitude and (unwrapped) phase are smooth functions of frequency, making them easy to interpolate. Evaluating this interpolant onto the FFT frequency grid with millions of points is still an expensive operation which is here accelerated by using a GPU. We use the \texttt{BBHx} GPU-based implementation of cubic-spline interpolation \citep{Katz:2020,Katz:2022}. 

The amplitude-phase decomposition is also a convenient representation to apply the LISA instrument response to the waveform. The LISA response can be represented using transfer functions which operate on each frequency mode and simulate the instruments response in the noise-orthogonal TDI channels $\alpha \in \{\mathrm{A},\mathrm{E},\mathrm{T}\}$ \citep{Armstrong:1999,Tinto:2021}.
\begin{equation}
    h_{\alpha}(\theta,f) = \mathcal{T}_{\alpha}(\theta,f)h(\theta,f)
\end{equation}
We use the \texttt{BBHx} implementation of the frequency domain LISA response, which performs interpolation for the transfer functions in a similar way \cite{Marsat:2018}. The sparse frequency grid over which both the response and waveform are evaluated, is a downsampled version of the FFT frequency grid, by a factor of 1000.

The formalism used to derive the response breaks down in the limit of extremely slowly chirping signals. However, for SmBBH sources the response breaks down at $t_c\gtrsim10-100$ years (see Fig. 3 from \cite{Marsat:2018}), all sources considered in this paper merge faster than this. It is expected that the large majority of SmBBH systems ($\sim 90\%$) merge with $t_c < 10 $years, see Fig. 3 of Ref.~\cite{Gerosa:2019}.

The computation of all quantities related to signal inner products that involve summation over the FFT frequency grid (including $x_n(d,\theta)$, $\Upsilon_N$ and $\mathrm{log}\hat{L}_N$) were also GPU accelerated, significantly reducing the computational cost of both search and inference. 

Throughout this work it has been assumed that the PSD that characterizes the noise spectrum in the data is known. 
The analytical approximation to the instrumental noise from Ref.~\cite{Babak:2021} was used. 
The confusion noise arising from the unresolved galactic DWD population was not included as it is expected to be negligible (i.e.\ smaller than the instrumental noise) at frequencies $f_{\rm{GW}}\gtrsim 10^{-2}\, \rm{Hz}$ \cite{Karnesis:2021} where all the sources considered in this paper reside. 
The noise is assumed to be Gaussian and is generated directly in the frequency domain from the known PSD. 
The simulated data stream has no gaps or glitches. 
Extending the search to data streams with these properties is left for future work.

The actual LISA noise spectrum is expected to drift in time and to contain significant cyclostationary components associated with orbital modulations of the confusion noise \cite{Digman:2022}. 
However, the confusion noise dominates at lower frequencies $f\lesssim 3\times 10^{-3}\,\mathrm{Hz}$; at the higher frequencies where the sources considered in this study reside, while the noise is not expected to be perfectly stationary, it will vary much less with time.

In this paper the search is performed on data sets containing signals from two different binaries (a GW150914-like and GW190521-like source \cite{Abbott:2016,Abbott:2020}) injected at multiple different SNRs (controlled by the choice of the $D_L$ parameter). 
Both sources are placed closer than the distance at which they were observed by the LVK in order to have an SNR in LISA which will be detectable. 
We also arbitrarily pick the time to merger (controlled by the choice of the $f_{\rm low}$ parameter) such that these systems coalesce within $\sim 4$ years (i.e.\ within the nominal LISA mission lifetime). 
The fiducial ``loud" source in this paper is similar to GW150914, placed at a very close distance of $50 \rm{Mpc}$. 
In Sec.~\ref{sec:search_results} we vary the distance to this source to evaluate the performance of this search at different SNRs. We simulate searches for signals at a range of SNRs, spanning the range $\rho \in [15,38]$, focusing more on the subset that span $\rho \in [15,22]$. While this SNR range is on the upper end of what LISA may observe, it is not extreme, see Fig. 6 of Ref.~\cite{Gerosa:2019}.

The injected source parameters and search priors on those parameters for the fiducial GW150914-like source are shown in Table \ref{tab:GW150914-like_injection}, the two parameters not directly searched over are the luminosity distance $D_L$ and the initial orbital phase $\phi_0$. A similar table for GW190521 is shown in Appendix \ref{app:GW190521}.
The luminosity distance adds a $1/D_L$ pre-factor to the waveform amplitude and is cancelled out of both $\tilde{\rho}$ and $\Upsilon_N$.
While in the definition of $\Upsilon_{N}$, the phase is analytically maximized so it does not need to be directly searched over. After the search has concluded and $\Upsilon_{N=1}$ has been optimized, in the process of automatically transitioning to inference, $D_L$ is drawn from the prior in Table \ref{tab:GW150914-like_injection} for each PSO particle which switches behavior to an MCMC walker. Since in the inference stage $\log\hat{L}_1$ is being sampled, $\phi_0$ is still maximized over.

We find empirically that this method is not able to search over the entire astrophysically reasonable parameter space for SmBBH signals at once. 
This is simply because, if the finite number of particles used here are spread too thinly across parameter space then the probability that a large peak is missed increases.
Swarms of particles must be initialized with sufficient density over parameter space to find a peak in the search statistic and climb it, thus the search space needs to be broken into multiple `tiles' each of which are searched independently. This can be done in parallel. We divide up the search space into rectangular tiles in the $\mathcal{M}_c$ and $t_c$ parameters. We have found empirically a tile dimension of $\sim3 \, \rm{M}_{\odot}$ and $6$ months to be effective. Tiling the astrophysically reasonable search space of $\mathcal{M}_c \in [10,100] \,  \rm{M}_{\odot}$ and $t_c \in [0.5,10]$ years using tiles of these dimensions would require around $\sim 600$ tiles. This is a rough estimate, and assumes each tile has similar dimensions, in reality for the high mass end of parameter space the tiles can be made bigger in both dimensions, since these undergo less orbital cycles. Fig.~\ref{fig:schematic_flowchart} shows a high level overview of the search process within one tile.

\begin{table}[t]
    \caption{
        Injection parameters and priors for the search of the GW150914-like fiducial source.
        Parameters in the top section of the table are those that are searched using uniform priors over the ranges shown, those in the middle section are injected but do not appear in the definition of the search statistic, and those in the bottom section are \emph{derived} parameters and approximate prior ranges are given. At the point of transition between search and parameter estimation, each particle draws $D_L$ uniformly from the prior range $[10,200] \, \rm{Mpc}$.
        All masses are given in the detector frame.
        (Due to differences in the conventions between the waveform and the BBHx LISA response, the definition of the polarisation angle $\psi$ acquires a negative sign.)
        Parameters that define the search tile are in highlighted rows. This injected source has an SNR of $38$.
    }
    \label{tab:GW150914-like_injection}
    \vspace{2pt}
    \begin{tabular}{c|cc}
        \hhline{|=|= =|}
        \hspace{0.33cm}Parameter\hspace{0.3cm} & \hspace{0.3cm}Injection\hspace{0.3cm} & \hspace{0.3cm}Prior range: $[\theta_{\rm min},\theta_{\rm max}]$\hspace{0.33cm} \\
        \hline
        \rowcolor{lightgray}$\mathcal{M}_c\,[M_\odot]$ & $28.095555$ & $[27, 30]$\\
        $f_{\rm{low}}\,[\mathrm{Hz}]$ & $0.018$ & $[0.0178, 0.0182]$ \\
        $\eta$ & $0.2471$ & $[0.15,0.2495]$ \\
        $\lambda\,[\mathrm{rad}]$ & $2.01$ & $[0,2\pi]$ \\
        $\beta\, [\mathrm{rad}]$ & $\pi/4$& $[-\pi/2,\pi/2]$ \\
        $i\, [\mathrm{rad}]$ & $2.498$& $[0,\pi]$ \\
        $\psi\, [\mathrm{rad}]$ & $-1.85$& $[-\pi,0]$ \\
        $e_0$ & $0.01$& $[0.005,0.1]$ \\
        \hline
        $\phi_0\, [\mathrm{rad}]$ & $0$& - \\
        $D_L\, [\mathrm{Mpc}]$ & $50$& - \\
        \hline
        $m_1\,[M_\odot]$ & $36$ & $\sim[32.44,76.43]$ \\
        $m_2\,[M_\odot]$ & $29$ & $\sim[15.49,32.96]$ \\
        \rowcolor{lightgray}$t_c \, [\rm{months}]$ & $43.28$ & $\sim [38,46]$\\ 
        \hhline{|=|= =|}
    \end{tabular}
\end{table}

\section{Statistical properties of the semi-coherent search statistic and likelihood}\label{sec:search_statistic_statistical_properties}

In this section the statistical properties of $\Upsilon_N$ and $\log\hat{L}_N$ are tested.
In Sec.~\ref{subsec:results_upsilon_dist} we validate the theoretical sampling distributions of $\Upsilon_N$ under both only noise and noise plus signal for various values of $N$. 
Then, in Sec.~\ref{subsec:PPplots}, the statistical properties of $\log\hat{L}_N$ are tested using probability-probability plots computed over many noise realizations.

\subsection{Sampling distributions of the semi-coherent search statistics} \label{subsec:results_upsilon_dist}

We seek to verify the distributions of the random variables $\Upsilon_N$ given in Sec.~\ref{sec:Search_statistics_variation_under_noise}, by computing $\Upsilon_N$ over many noise realizations and comparing the empirical distributions against those derived from the CLT (see Eqs.~\ref{eqn:CLTdist_noise} and \ref{eqn:CLTdist_signal}). We compare the histograms of these empirically obtained random variables against the theoretical probability density function. The GW source used for all computations in this subsection is the GW150914-like fiducial source in Table \ref{tab:GW150914-like_injection}. The luminosity distance to the source is scaled to simulate signals with $\rho \sim \{10,20,30\}$. 

\begin{figure}[!t]
    \centering
    \includegraphics[width=0.46\textwidth]{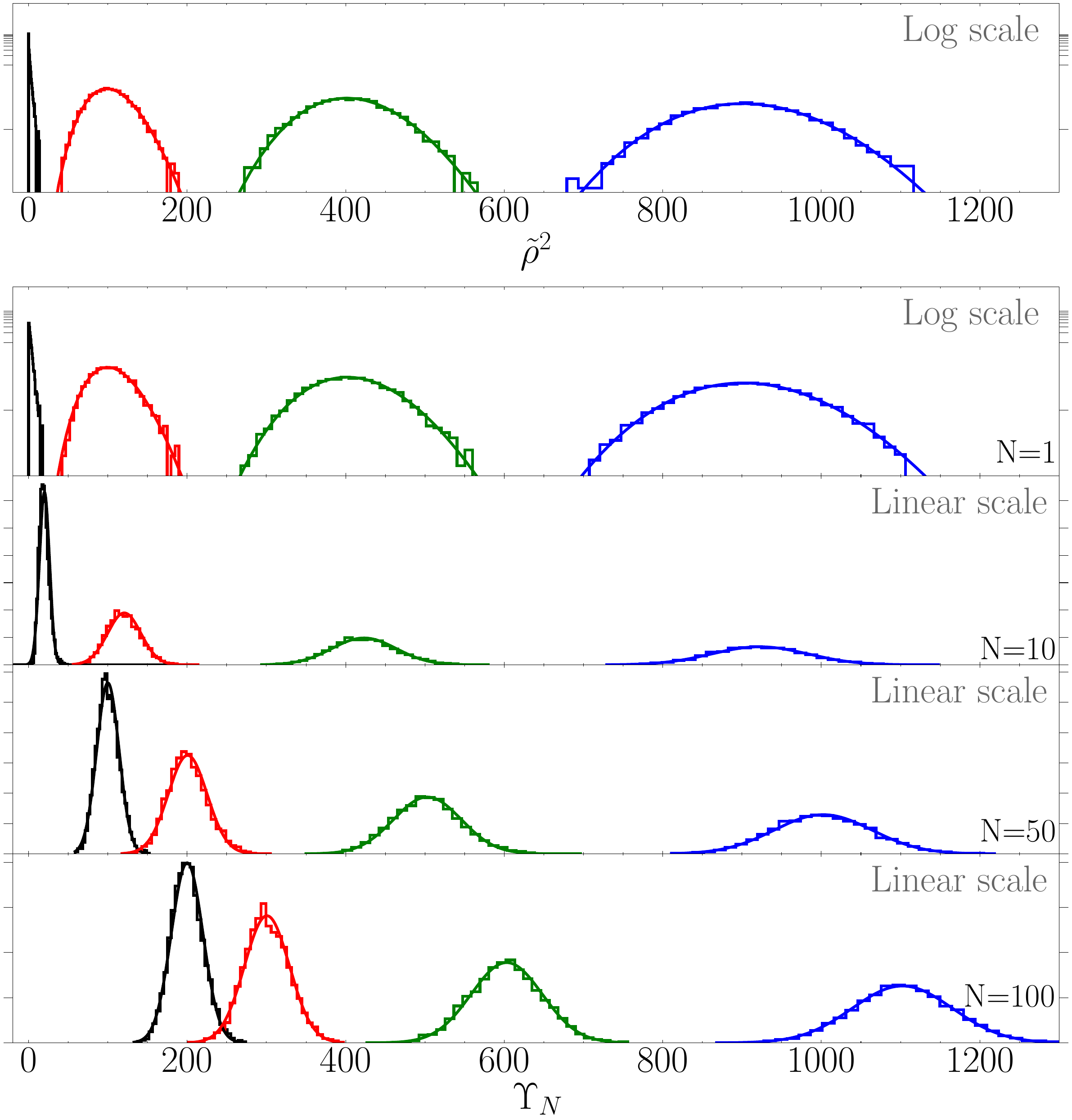}
    \caption{Distributions of the coherent $\tilde{\rho}^2$ (\textit{Top panel}) and semi-coherent $\Upsilon_N$ (\textit{Bottom panel}) matched filter statistics, sampled in the noise only ($d=n$; black) and signal ($d=h$) cases, and for varying injected SNRs: $10$ (red), $20$ (green) and $30$ (blue).
    Each vertical panel represents a different number of segments $N$. The top panel plots $\tilde{\rho}^2$, instead of simply $\tilde{\rho}$, as this quantity is more comparable with $\Upsilon_1$. The plots for the coherent $\tilde{\rho}^2$ and $\Upsilon_1$ (top two plots) use a log-scaled $y$-axis while the lower three plots use a linear scale. The signal used for injections in this plot is the GW150914-like loud fiducial source from table \ref{tab:GW150914-like_injection}.
    The analytic predictions for the distributions are plotted using smooth lines (except for the $\Upsilon_{N=1}$ and $\tilde{\rho}^2$ cases as they diverge at $0$). In all cases $5000$ independent noise realizations were used to construct the histograms.} 
    \label{fig:False_alarm_prob}
\end{figure}

The well known Gaussian sampling distributions for $\tilde{\rho}$ are checked in the top panel of Fig. \ref{fig:False_alarm_prob}. In the noise only case $\tilde{\rho} \sim \mathcal{N}(0,1)$, whereas in the noise+signal case, $\tilde{\rho} \sim \mathcal{N}(\rho,1)$ where $\rho$ is the optimal SNR of the signal. In Fig.~\ref{fig:False_alarm_prob} we plot $\tilde{\rho}^2$ since it is directly comparable with $\Upsilon_1$. In the noise only case $\tilde{\rho}^2 \sim \chi_1^2$, a standard chi squared distribution with 1 degree of freedom. While in the noise+signal case $\tilde{\rho}^2 \sim \chi_1^2(\lambda,1)$, a non-central chi squared distribution with non-centrality parameter $\lambda = \rho^2$ and unit scale parameter. 
The bottom panel of Fig~\ref{fig:False_alarm_prob} verifies the distributions given in Sec.~\ref{sec:Search_statistics_variation_under_noise} for $\Upsilon_N$ at $N \in \{1,10,50,100\}$. The analytic approximations from the CLT are used as the theoretical distribution for $N\geq10$, while for the $\Upsilon_1$ case these are the theoretical $\chi^2$ distributions derived in the same section. 
The increased FAP of $\Upsilon_N$ as a function of $N$ can be seen in Fig. \ref{fig:False_alarm_prob} as the overlap between the two distributions $\Upsilon_N(d=h+n,\theta)$ and $\Upsilon_N(d=n,\theta)$ increases with $N$. This is most clearly seen when comparing the black and red curves. 
The distributions of $\Upsilon_1$ and $\tilde{\rho}^2$ are similar, indicating that either can be used as a good detection statistic.
In our search, the significance of a trigger is determined at the final stage, when $N=1$, using this statistic. 

\subsection{PP-type plots for the semi-coherent likelihood} \label{subsec:PPplots}

\begin{figure}[!ht]
    \centering
    \includegraphics[width=0.48\textwidth]{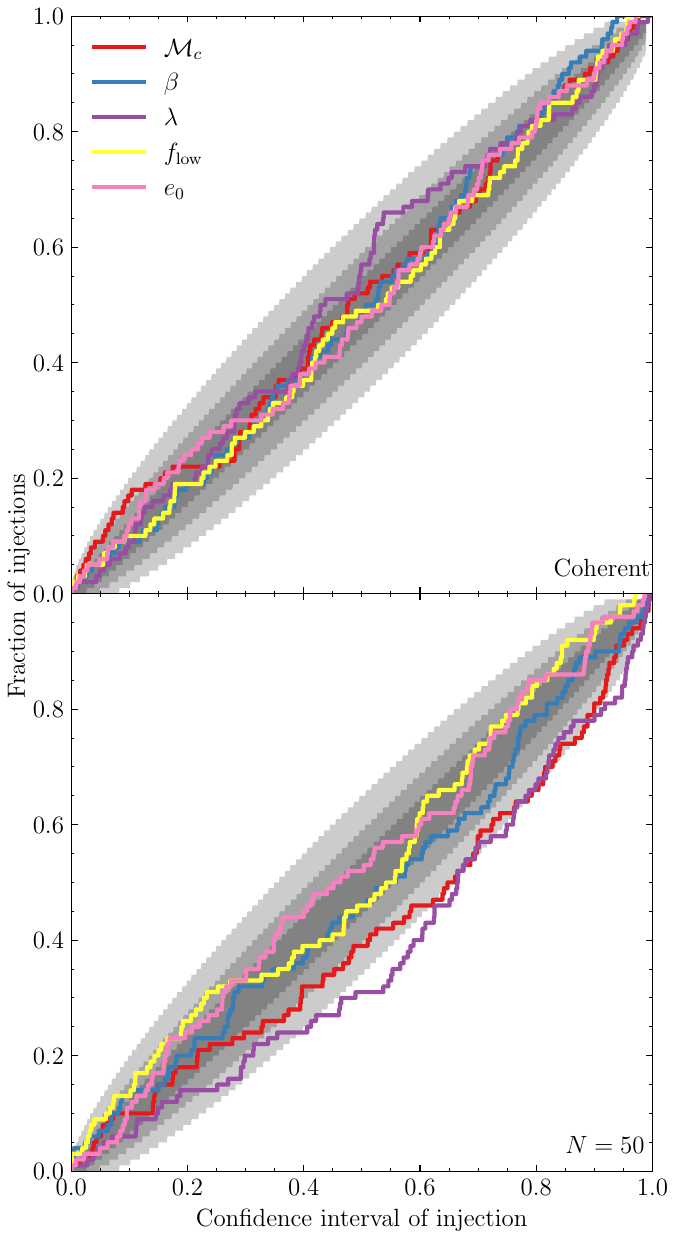}
    \caption{Probability-probability plot for the coherent (top panel) and $N=50$ semi-coherent likelihoods (bottom panel), computed over $100$ injection-inference runs of the GW150914-like source into simulated Gaussian noise. Shaded regions denote the $1, \, 2$ and $3\sigma$ confidence intervals. The $5$ colored curves show the 1-dimensional PP plots for the key phasing parameters in the legend. PP plots are shown for chirp mass $(\mathcal{M}_c)$, ecliptic latitude $(\beta)$, ecliptic longitude $(\lambda)$, initial orbital frequency $(f_{\rm{low}})$ and initial eccentricity $e_0$.}
    \label{fig:Combined_PP_subset}
\end{figure}

Probability-probability (PP) plots are a common statistical test in GW astronomy used to verify many ingredients in parameter estimation pipelines, including the likelihood function (see, for example, Ref.~\cite{Veitch:2015}).
Here, this is used to check the properties of the semi-coherent likelihood $\log \hat{L}_{N}$.
Peaks of the semi-coherent likelihood are (by design) wider than corresponding peaks in the usual (coherent) likelihood, this makes them easier to find in a search.
It is also expected that the semi-coherent likelihood is more susceptible to noise fluctuations and that the width of the peaks should be consistent with this increased scatter.
If this is true, then the PP plot should remain approximately diagonal.

Each data-point in the PP plot corresponds to one full parameter estimation run conducted with either a coherent or semi-coherent likelihood. The fiducial GW150914-like signal is used for all parameter estimation runs, each data-stream contains an independent realization of the noise. 
Parameter estimation is performed on each data-stream with narrow priors using the \texttt{dynesty} nested sampler \cite{Speagle:2020}. The 1D confidence interval the injection resides on is defined as
\begin{equation}
    X = \int\limits_{\{x|P(x)>P(x_*)\}} P(x) \mathrm{d}x .
\end{equation}
Where $P(x)$ is the one-dimensional marginalized posterior distribution on a parameter $x$ and $x_*$ corresponds to the injected value of that parameter. The discrete posterior samples from inference were fit with a Kernel Density Estimate (KDE), which provides a continuous approximation for $P(x)$. Then both $P(x_*)$ and $P(x_i)$ (posterior density for every posterior sample), can be estimated with the KDE. The integral required to compute the confidence level can then be approximated as a Monte-Carlo sum over the posterior samples:
\begin{equation}
    X \approx \frac{1}{\hat{N}} \sum_{i=1}^N \mathbf {1}_{\{x|P(x_i)>P(x_*)\}} (x_i)
 \end{equation}

Where $\mathbf{1}$ denotes the indicator function and $\hat{N}$ is the total number of posterior samples. This process is repeated 100 times to construct the PP plot. Finally, the fraction of injections that lie at various confidence levels are plotted on Fig.~\ref{fig:Combined_PP_subset}. The figure shows these results for two cases: a coherent and semi-coherent $N=50$ posterior. 

The parameter estimation runs used to populate Fig.~\ref{fig:Combined_PP_subset} used the \texttt{dynesty} nested sampler instead of the MCMC sampler \texttt{Zeus},  this is because the automated parameter estimation can be seeded directly from the search, which provides good starting locations for the MCMC walkers, close to the maxima.
Dynesty is only used to explore the likelihoods in this section, it is not part of the search pipeline that is the main focus of this paper.

We find that the diagonal property does approximately hold, see Fig.~\ref{fig:Combined_PP_subset}. The PP plots based on the semi-coherent likelihoods (e.g.\ $N=50$) do exhibit a small ``sag''. We conjecture that this is due to the maximization (as opposed to marginalization) over the phase angles in each segment. However, this is a small effect (quantified by a maximum sag of $\sim 10\%$, see Fig.~\ref{fig:Combined_PP_subset}). 
But this effect is small enough so that our search is still able to track peaks accurately through the different hierarchical stages of the search.

It is worth highlighting the PP plots in Fig.~\ref{fig:Combined_PP_subset} use only one source for injections (the GW150914-like source in Table \ref{tab:GW150914-like_injection}) but with multiple noise realizations. Usually, PP plots should be generated with many sources drawn randomly from the prior. For computational reasons, here we use a single source at a fixed $\theta$, this allows us to test the algorithm at a reasonable computational cost.

\section{Search results}\label{sec:search_results}

\begin{figure*}[ht!]
    \centering
    \includegraphics[width=\textwidth]{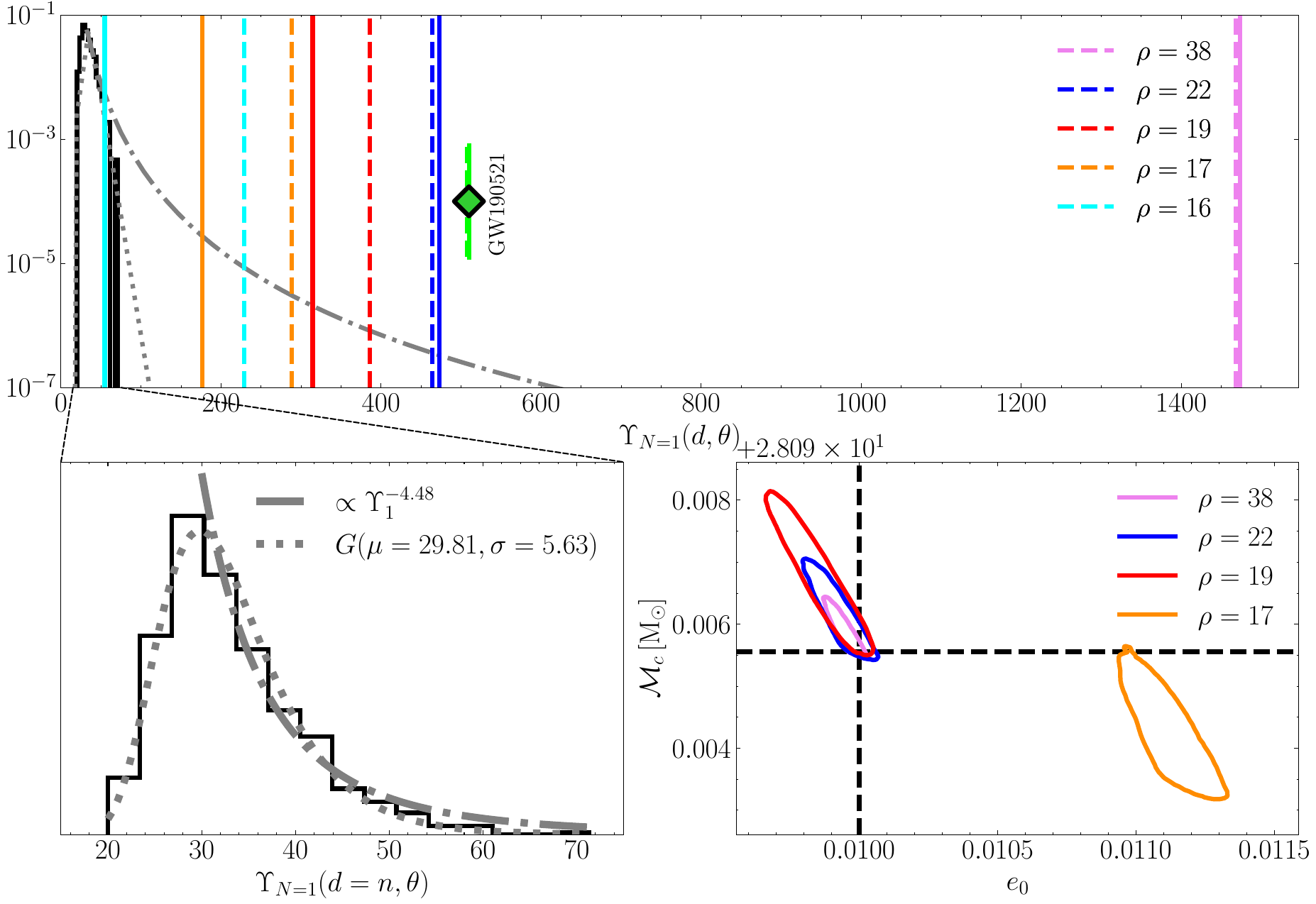}
    \caption{
    \textit{Top:} The search result is the maximum value of the $\Upsilon_1$ statistic and this is shown for each GW150914-like source (solid vertical colored lines).
    Also shown for comparison is the value of the search statistic evaluated at the injected parameters (dashed vertical colored lines).
    Also plotted is the background distribution of triggers in noise-only data (black histogram). 
    The search result for a GW190521-like event is also shown as a short line, although this comes from a search in a very different region of parameter space and so this result is only intended to be qualitative, for the reasons explained below.
    Gray curves show extrapolations of the background noise distribution and can be used to estimate the FAP. Two such extrapolations are shown, these are intended to bracket the optimistic and pessimistic cases, and these are also shown in the bottom left panel.
    \textit{Bottom left:} Distribution of background noise triggers, this is the same as the black distribution plotted in the top panel. This histogram is constructed from 60 noise only data streams. The dash-dotted gray line indicates a conservative power-law fit to the upper tail of this distribution with PDF $\propto \Upsilon_{N=1}^{-4.48}$, while the dotted line indicates a fitted Gumbel distribution which decays faster than the power law and provides a more optimistic extrapolation of the distribution of background triggers at high $\Upsilon_1$.
    This noise distribution was calculated from the search tile containing the GW150914-like source and therefore does not apply to the GW190521-like event.
    \textit{Bottom right:} $90$th percentile confidence contours on the chirp mass and eccentricity from the final MCMC stage of the search (not including the un-recovered $\rho=16$ case).  
    The characteristic lopsidedness of the posterior distribution is due to the degeneracy between $\mathcal{M}_c$ and $\eta$ and the parameter space boundary at $\eta=0.25$ (similar to that shown in Fig.~8 of \cite{Buscicchio:2021}). 
    For SNRs $\rho\geq 19$ the final MCMC stage of the search finds the true parameters within the $90$th percentile confidence contour. However at the lower SNR $\rho=17$ a significant bias is observed, this is because this is a crude MCMC, and is only intended to be rough estimate for the posterior on the source parameters, it is only intended to seed a more accurate parameter estimation.}
    \label{fig:real_false_alarm_prob_combined_figure}
\end{figure*}

This section presents the results of a search for a number of GW150914-like sources injected at five SNRs (from 16 to 38, see legend of Fig.~\ref{fig:real_false_alarm_prob_combined_figure}) into mock LISA data.
We describe how to quantify the significance of detections by estimating the FAP. 
Additionally, in Appendix \ref{app:GW190521} similar results are presented for the search of a GW190521-like source.

As described above, the search uses particle swarms to track multiple peaks in the semi-coherent search statistic $\Upsilon_N$.
The search is hierarchical in the sense that the number of segments $N$ used is steadily decreased throughout the search. 
At the final coherent stage ($N=1$), when the PSO algorithm converges to a number of peaks, this process stops.  
In a typical case, the number of peaks found by this process is $\sim 10$. However, many of these peaks do not correspond to real signals. The parameter values of these peaks are the candidate sources produced by the search.
For each of our searches, the highest of these peaks (i.e.\ the one with the largest value of $\Upsilon_{N=1}$) is indicated by a solid vertical line in the top panel of Fig.~\ref{fig:real_false_alarm_prob_combined_figure}. 

If the injected signal is sufficiently loud, then the search succeeds in locating the source in parameter space.
This can be seen from the results in Fig.~\ref{fig:real_false_alarm_prob_combined_figure}.
The top panel shows the highest value of the search statistic found by the particle swarms compared to the value of that statistic evaluated at the injected parameters; the optimized value is generally slightly higher due to noise fluctuations, similar to how the peak in the coherent log-likelihood is shifted from the true parameters when noise is present. 

For SNRs $\rho \geq 19$ the search finds the correct source parameters.
For SNRs $17 \leq \rho < 19$ the search often fails to find the exact peak in $\Upsilon_1(d,\theta)$, but it gets extremely close. In these cases the optimized value of the search statistic is less than the value at the injected parameters, and the parameters are somewhat offset from their true values.
For lower SNRs $\rho \leq 16$ the search does not find the source. 

\begin{figure*}[t!]
    \centering
    \includegraphics[width=0.85\textwidth]{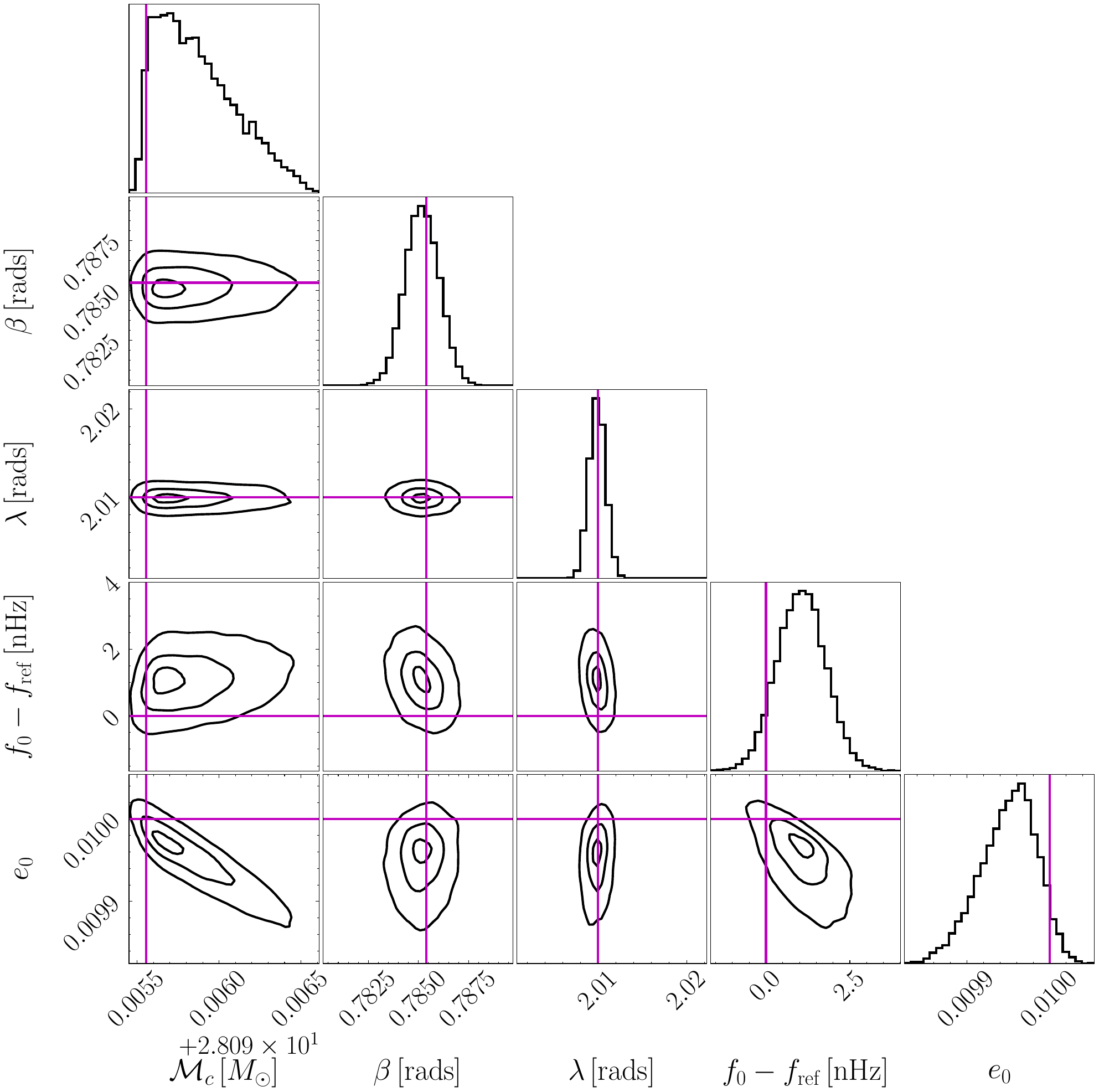}
    \caption{Parameter estimation results from the search for the GW150914-like fiducial source in Table \ref{tab:GW150914-like_injection}. Plotted parameters here are: chirp mass $(\mathcal{M}_c)$, ecliptic latitude $(\beta)$, ecliptic longitude $(\lambda)$, initial orbital frequency $(f_{0})$ and initial eccentricity ($e_0$). The $[10,50,90]$ percentile contours are plotted in black and the injected source parameters are marked by the magenta lines. Posteriors in the initial orbital frequency are plotted with respect to the reference frequency $f_{\rm{ref}} = 9\times10^{-3}\, \rm{Hz}$, which is also the injected value.}
    \label{fig:GW150914_SNR_38_PE}
\end{figure*}

It is necessary to assign a FAP to each of candidates found by the search.
The FAP is determined by comparing the maximum optimized value of the $\Upsilon_{N=1}$ statistic found by the search to the noise background distribution.
Unlike in the LVK case, where there are multiple independent instruments and large fractions of signal-free data, it is not possible to perform time slides of LISA data to determine this background. 
Instead, the background is computed by an injection campaign in which the search is performed on a large number of signal-free injections. 
These injections should ideally contain realistic LISA data but with no SmBBH sources present; here, for simplicity, 60 such searches were performed on time series containing only stationary, Gaussian noise.
Exactly the same hierarchical, semi-coherent PSO search algorithm was applied to each of these signal-free injections and the peak values of the $\Upsilon_{N=1}$ statistics were recorded.
This defines the noise background distribution of our search statistic and is plotted as a black histogram in the top and bottom left panels of Fig.~\ref{fig:real_false_alarm_prob_combined_figure}.

The FAP for a given candidate with maximum optimized value of $\Upsilon^*_{N=1}$ is defined as the complimentary cumulative density function of the noise background evaluated at $\Upsilon^*_{N=1}$. 
Similar techniques are used to determine the significance of candidates in LVK searches \cite{Cabourn:2022,Messick:2017}.

Given the noise background of 60 trials shown in Fig.~\ref{fig:real_false_alarm_prob_combined_figure} it is possible to bound the FAP of the candidates in the $\rho\geq 17$ searches as $\mathrm{FAP}\lesssim 1/60$.
In order to obtain better estimates of the FAP the background distribution needs to be extrapolated.
This is done in two ways: first, a power law model was fit to the noise background above $\Upsilon_{N=1}\geq 30$, and secondly a Gumbel distribution was fitted to the entire background distribution. 
The power law represents a conservative extrapolation, the real distribution of triggers is expected to decay faster than this, see for example Fig.~4 of Ref.~\cite{Abbott:2016}. 
The Gumbel \cite{Tenorio:2022} distribution is more optimistic and decays faster (exponentially). 
The motivation for using the Gumbel distribution is that in the ideal case of Gaussian noise, $\Upsilon_{N=1}$ at a particular point in parameter space on the signal-free injections is known to be distributed as a $\chi^2$ statistic. The maximum of many draws from a $\chi^2$ distribution approaches the Gumbel distribution \cite{Tenorio:2022}. These two extrapolations are used to compute the FAP presented in Table \ref{tab:false_alarm}.

\begin{table}[t!]
    \caption{
    Estimates for the FAP of the searches for the different GW150914-like sources. In each case two estimates are provided, one using the power-law extrapolation of the noise background (conservative, i.e.\ higher FAP) and one using the Gumbel extrapolation (optimistic, i.e.\ lower FAP).
    }
    \label{tab:false_alarm}
    \vspace{2pt}
    \begin{tabular}{c|cc}
        \hhline{|=|==|}
        SNR $\rho$ & FAP (lower)  &  FAP (upper) \\
        \hline
        16 & $7.4 \times 10^{-2}$ & $1.2 \times 10^{-2}$\\
        17 & $1.3 \times 10^{-3}$ & $5.1 \times 10^{-12}$\\
        19 & $1.6 \times 10^{-4}$ & $< 10^{-12}$\\
        22 & $4.1 \times 10^{-5}$ & $< 10^{-12}$\\
        38 & $7.8 \times 10^{-7}$ & $< 10^{-12}$\\
        \hhline{|=|==|}
    \end{tabular}
\end{table}

The FAP numbers provided should just be considered as rough estimates. In addition to the extrapolation from a fairly small sample from the background distribution, computing an accurate FAP requires simulation of realistic signal-free injections. In this study, these were highly simplified in that they do not contain gaps, any other source types, glitches and the noise is stationary. 
However, the results of this study suggest that signals with $\rho \geq 20$ seem to be reliably detectable with this search pipeline while those at $\rho \leq 16$ seem to be completely undetectable. This is roughly consistent with the detection probability of $\Upsilon_{N=100}$, which is expected to decay from $1$ to $0$ in the range $\rho \in [15,20]$, see Fig.~7 in Ref.~\cite{Chua:2017}. 
For intermediate SNRs, where the source is only detected some of the time, this detection probability is explored by performing multiple simulated searches with the results presented in Appendix \ref{app:Many_searches}.

Once a significant candidate has been found, the final stage of the search involves an MCMC (see Sec.~\ref{subsec:automated_PE}).
This changes the behavior of the swarm of particles from optimizing and finding the peak of $\Upsilon_{N=1}$ to exploring, or sampling, the posterior distribution based on $\log \hat{L}_{N=1}$.
It should be stressed that this pipeline is not intended to perform high-quality sampling of the posterior.
Because the MCMC walkers are all seeded near the peak found by the PSO the subsequent sampling will be slow to properly explore the tails of the posterior and will likely miss entirely any secondary peaks.
However, this relatively crude MCMC can still be useful because it gives an indication of the posterior width in each parameter and can be used to seed more detailed follow-up inference.

These parameter estimation results from the search of the GW150914-like $\rho=38$ fiducial source are shown in Fig.~\ref{fig:GW150914_SNR_38_PE}. 
Additionally, posteriors on the chirp mass and eccentricity parameters are shown for all recovered sources in the bottom right panel of Fig.~\ref{fig:real_false_alarm_prob_combined_figure}.
Posteriors are only shown for the phasing parameters; i.e.\ parameters that significantly contribute to the GW phase. 
This final MCMC stage of the search performs well for loud sources, where the PSO reliably and accurately locates the peak. For quieter sources, such as $\rho=17$ case, the parameter estimation results are biased even though this source is confidently recovered.

\section{Computational cost}\label{sec:computational_cost}

The search described in this paper is designed to be broken down into distinct tiles in $\mathcal{M}_c -t_c$ space. Each tile can be searched over simultaneously and we have shown example searches within a couple of these tiles. Each search in this paper has used 1 NVIDIA A100 GPU and 10 CPUs, we have made use of the \texttt{PyTorch} implementation of \texttt{multiprocessing} to parallelize the process of computing waveforms and search statistics on one GPU \cite{Paszke:2019}. All the searches presented required $\sim 1-2$ days of compute time. This does not include the parameter estimation described in Sec.~\ref{subsec:automated_PE}. For the small number of sources where this parameter estimation was performed this took an additional $\sim 10$ hours per candidate. Given that $\sim 600$ tiles will be needed to cover the parameter space of interest and the expected small number of sources in LISA, while expensive, this search can be scaled up the required level. Additionally future hardware and algorithmic improvements will reduce the cost further. 

\section{Conclusion}\label{sec:discussion}

In this paper we have demonstrated a GPU accelerated semi-coherent hierarchical search for SmBBH inspirals in simulated LISA data containing stationary Gaussian noise. We have established the threshold SNR for this search to be around $\rho_{\rm{threshold}} \approx 17$. This is roughly consistent with previous estimates of $\rho_{\rm{threshold}}=15$ for this threshold \cite{Moore:2019}. Improvements to the optimization and search algorithm may push the threshold SNR down slightly further. We have also demonstrated a method for computing the background distribution of noise triggers from a stochastic search, and used it to place upper and lower limits on the FAP for triggers arising from a few simulated searches. Realistic simulated LISA data is required to make robust estimate of the FAP and detection probability for this type of search.

The work done in this paper is limited to the sources with low eccentricity and no spin, however we have some promising early results that indicate the semi-coherent search should work well on aligned-spin sources. In addition to including spin, the main focuses of future work will be to extend this method to data-streams with non-stationary components, gaps and glitches, and also to search the LISA Data Challenge datasets which contain SmBBH signals within them. We will also explore lowering $N_{\rm{high}}$ and analytically maximizing over more of the extrinsic search parameters, as this may allow for a lower threshold SNR as demonstrated in Ref.~\cite{Fu:2024}.

\vspace{0.4cm}
\begin{acknowledgments}
We would like to thank Rodrigo Tenorio for very helpful discussion about the false alarm probability relating to continuous wave searches. We would also like to thank Gonzalo Morrás for useful discussions on semi-coherent continuous wave methods. The computations described in this paper were performed using the University of Birmingham’s BlueBEAR HPC service. DB is supported by the STFC. DB also acknowledges the support of the Google Cloud Research Credits program for the early prototyping of this project, Grant No. 289387648. CJM acknowledges the support of the UK Space Agency, grant No. ST/V002813/1. This paper made use of \texttt{CuPy} and \texttt{Numba} for various GPU/CPU accelerated functionalities \cite{Okuta:2017,Lam:2015}.
\end{acknowledgments}

\section{Code availability}
The library used to perform this search can be found at \href{https://github.com/dig07/SC_Search}{\url{https://github.com/dig07/SC_Search}}. This library is currently still under early development, however the parts of the code relevant to the semi-coherent method can be easily reproduced from \href{https://github.com/dig07/SC_Search/tree/main/SC_search/Semi_Coherent_Functions.py}{\url{https://github.com/dig07/SC_Search/tree/main/SC_search/Semi_Coherent_Functions.py}}. While all the results in this paper can be reproduced using this code, its not designed to be used by itself and will in future be implemented in a larger library. 


\bibliographystyle{apsrev4-1}
\bibliography{bibliography}

\newpage
\appendix

\begin{table}[hbt!]
    \caption{
        Injection parameters and priors for the search of the GW190521-like fiducial source.
        Parameters in the top section of the table are those that are searched using uniform priors over the ranges shown, those in the middle section are injected but do not appear in the definition of the search statistic, and those in the bottom section are \emph{derived} parameters and approximate prior ranges are given. At the point of transition between search and parameter estimation, each particle draws $D_L$ uniformly from the prior range $[10,300] \, \rm{Mpc}$.
        All masses are given in the detector frame.
        (Due to differences in the conventions between the waveform and the BBHx LISA response, the definition of the polarisation angle $\psi$ acquires a negative sign.)
        Parameters that define the search tile are in highlighted rows. This source has an SNR of $21.45$.
    }
    \label{tab:GW190521-like_injection}
    \vspace{2pt}
    \begin{tabular}{c|cc}
        \hhline{|=|= =|}
        \hspace{0.33cm}Parameter\hspace{0.3cm} & \hspace{0.3cm}Injection\hspace{0.3cm} & \hspace{0.3cm}Prior range: $[\theta_{\rm min},\theta_{\rm max}]$\hspace{0.33cm} \\
        \hline
        \rowcolor{lightgray}$\mathcal{M}_c\,[M_\odot]$ & $95.0209$ & $[93, 96]$\\
        $f_{\rm{low}}\,[\mathrm{Hz}]$ & $0.0175$ & $[0.014, 0.019]$\\
        $\eta$ & $0.234$ & $[0.15,0.2495]$\\
        $\lambda\,[\mathrm{rad}]$ & $3.24$ & $[0,2\pi]$\\
        $\beta\, [\mathrm{rad}]$ & $0.4$& $[-\pi/2,\pi/2]$\\
        $i\, [\mathrm{rad}]$ & $2.0$& $[0,\pi]$\\
        $\psi\, [\mathrm{rad}]$ & $-1.5$& $[-\pi,0]$\\
        $e_0$ & $0.03$& $[0.005,0.1]$\\
        \hline
        $\phi_0\, [\mathrm{rad}]$ & $0$& - \\
        $D_L\, [\mathrm{Mpc}]$ & $200$& - \\
        \hline
        $m_1\,[M_\odot]$ & $142$ & $[111.74,244.58]$\\
        $m_2\,[M_\odot]$ & $85$ & $[53.35,105.47]$\\
        \rowcolor{lightgray}$t_c \, [\rm{months}]$ & $6.12$ & $\sim [5,11]$\\ 
        \hhline{|=|= =|}
    \end{tabular}
\end{table}

\section{Linearly chirping signal}\label{app:linear_chirp}

Throughout this paper, when signals and templates are split into semi-coherent segments, this splitting is done in the frequency domain such that each segment contains equal squared SNR.
As discussed in the main text, for a simple source this can be related to splitting into equal duration segments in the time domain. This appendix demonstrates this for a simple source. 

Consider a linearly chirping source with a constant amplitude,
\begin{equation}
    h(t) = \cos(2\pi\bigg[f t + \frac{\dot{f}}{2}t^2\bigg]).
\end{equation}
If the source is quasi-monochromatic (i.e.\ only increases its frequency in a narrowband across the observation period), $\dot{f}_{\rm GW} \ll f_{\rm GW}/T_{\rm{obs}}$, and the PSD can be approximated as constant across this narrow bandwidth (i.e.\ white noise), then the SNR can be computed in the time domain (Parseval's theorem). 
The cumulative square SNR up to a given time $t$ is given by
\begin{align} \label{eq:sq_snr_linear}
    \rho^{2}(t) &\propto \int_0^{t} \cos^2\bigg(2\pi\bigg[f t' + \frac{\dot{f}}{2}t'^2\bigg]\bigg) \mathrm{d}t' \nonumber \\
    &\propto \frac{1}{2}\int_0^{t} \bigg[\cos\bigg(4\pi\bigg[f t' + \frac{\dot{f}}{2}t'^2\bigg]\bigg)+1\bigg] \mathrm{d}t' \nonumber \\
    &\propto \frac{t}{2} + \frac{1}{2}\int_0^{t} \cos\bigg(4\pi\bigg[f t' + \frac{\dot{f}}{2}t'^2\bigg]\bigg) \mathrm{d}t'.
\end{align}
The second term evaluates to 
\begin{align}
    \frac{1}{2\sqrt{\dot{f}}}&\bigg( \cos(\frac{2 f^2 \pi}{\dot{f}})\bigg[-\mathcal{C}\bigg(\frac{2f}{\sqrt{\dot{f}}}\bigg)+\mathcal{C}\bigg(\frac{2(f+\dot{f}t)}{\dot{f}}\bigg)\bigg] + \nonumber \\
    &\sin(\frac{2 f^2 \pi}{\dot{f}})\bigg[-\mathcal{S}\bigg(\frac{2f}{\sqrt{\dot{f}}}\bigg)+\mathcal{S}\bigg(\frac{2(f+\dot{f}t)}{\dot{f}}\bigg)   \bigg]\bigg).
\end{align}
Where $\mathcal{C}$ and $\mathcal{S}$ are the fresnel integrals. 

\begin{figure}[!h]
    \centering
    \includegraphics[width=0.46\textwidth]{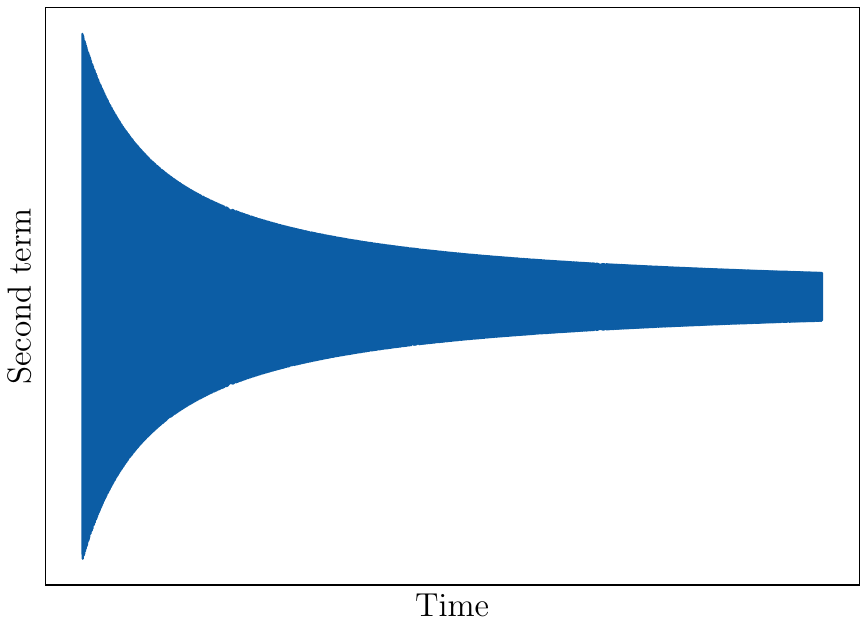}
    \caption{Sketch showing the behaviour of the second term in the integral.}
    \label{fig:Fresnel_integral}
\end{figure}

Fig.~\ref{fig:Fresnel_integral} sketches this as a function of time. This is a highly oscillatory function that decays to $0$ as $t \rightarrow \infty$. 
Therefore, for large observation times this term in negligible and the squared SNR in Eq.~\ref{eq:sq_snr_linear} increases linearly with time.

\section{PSO \& Inference parameters}\label{app:PSO_params}
In this appendix we summarise the parameters used for the PSO algorithm used in this paper.
The search is initialized with 6 swarms, each with 10,000 particles and the total number of particles is kept constant throughout the search. All of the searches conducted in this paper use the segment ladder $N=\{100,50,10,1\}$. We have used $N_{\rm{high}}=100$ as this was estimated to be the minimum number of segments required for a search in Ref.~\cite{Gair:2004} (this was computed for EMRIs, however the arguments still hold for SmBBH searches). A swarm is considered converged if the best value of the objective function in the history of the swarm does not improve by more than 2 in 50 iterations.

The minimum velocities in each dimension for each segment in the ladder are presented in Table \ref{tab:PSO_hyperparams}, alongside the $3$ PSO hyper-parameters.

At the end of the optimization phase of the search, if a swarm passes the noise threshold, 100 PSO particles transition to MCMC walkers, evolving for 10,0000 iterations.

\section{GW190521-like search}\label{app:GW190521}

\begin{figure}[h!]
    \centering
    \includegraphics[width=0.5\textwidth]{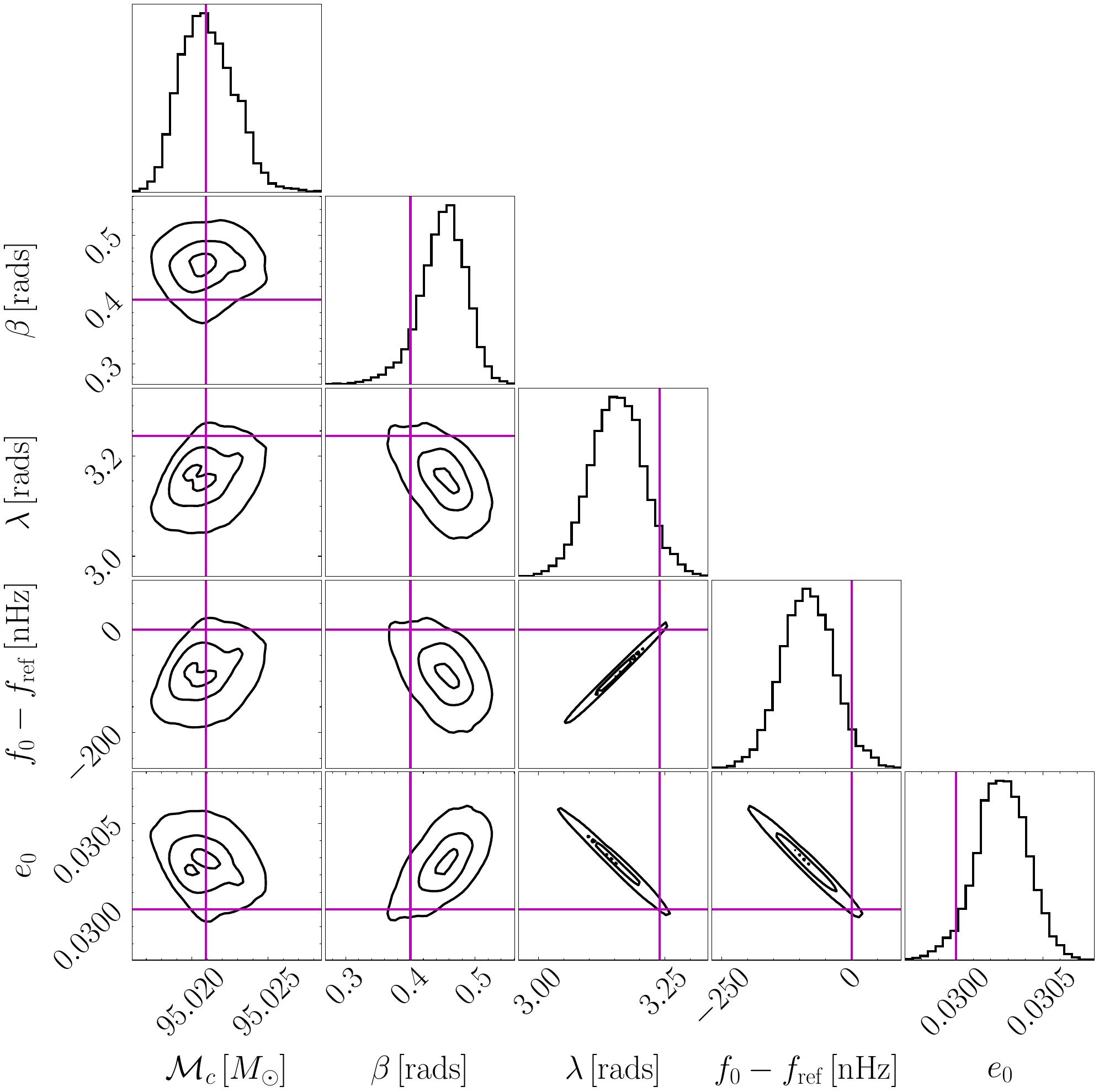}
    \caption{Parameter estimation result from the search of the GW190521-like source with source parameters shown in Table \ref{tab:GW190521-like_injection}. Posteriors are plotted for the same parameters as in Fig.~\ref{fig:GW150914_SNR_38_PE}. For this source, $f_{\rm{ref}} = 8.75\times10^{-3}\, \rm{Hz}$.}
    \label{fig:GW190521_SNR_21_PE}
\end{figure}

This section contains results for a simulated search for a GW190521-like signal. The injected parameters and priors are shown in Table \ref{tab:GW190521-like_injection}. The results of the search are presented in Figs.~\ref{fig:real_false_alarm_prob_combined_figure} and \ref{fig:GW190521_SNR_21_PE}. For this search it is not possible to estimate the FAP because we did not perform a signal-free injection and search campaign in this search tile.

\begin{table*}
\caption{PSO hyper-parameters and minimum velocities used throughout the search.}
\begin{ruledtabular}
\begin{tabular}{c|ccc|cccccccc}
    \textrm{Number of segments (N)} &
    \textrm{$\Omega$} & 
    \textrm{$\Phi_P$} & 
    \textrm{$\Phi_G$} &
    \textrm{$\epsilon^{\mathcal{M}_c}$} & 
    \textrm{$\epsilon^{\eta}$} & 
    \textrm{$\epsilon^{\beta}$} & 
    \textrm{$\epsilon^{\lambda}$} & 
    \textrm{$\epsilon^{i}$} & 
    \textrm{$\epsilon^{\psi}$} & 
    \textrm{$\epsilon^{f_{\rm{low}}}$} & 
    \textrm{$\epsilon^{e_0}$} \\
    \colrule
    $100$ & $0.5$ & $0.2$ & $0.3$ & $0.1$ & $0.05$ & $0.1$ & $0.1$ & $0.2$ & $0.1$ & $5\times10^{-8}$ & $0.1$ \\
    $50$ & $0.5$ & $0.2$ & $0.4$ & $0.01$ & $0.05$ & $0.01$ & $0.01$ & $0.2$ & $0.1$ & $1\times10^{-8}$ & $0.1$ \\
    $10$ & $0.3$ & $0.2$ & $0.5$ & $0.001$ & $0.05$ & $0.001$ & $0.001$ & $0.01$ & $0.1$ & $1\times10^{-8}$ & $0.1$ \\
    $1$ & $0.72$ & $1.193$ & $1.193$ & $0.0$ & $0.0$ & $0.0$ & $0.0$ & $0.0$ & $0.0$ & $0$ & $0.0$ 
    \label{tab:PSO_hyperparams}
\end{tabular}
\end{ruledtabular}
\end{table*}

\section{Detection probability}\label{app:Many_searches}

For intermediate SNRs, the search only correctly identifies the signal some of the time. This detection probability is a function of SNR. This can be assessed by searching multiple injections at the same SNR in independent noise realizations.
In this section we make a limited attempt of this by simulating 4 searches for signals at a given SNR, each with its own realization of the instrumental noise. This process is repeated across a range of SNRs. The results of this can be seen in Tables.~\ref{tab:Best_upsilons} and \ref{tab:Best_matches}.
Based on these results the search seems to reliably find sources with $\rho \gtrsim 20$. For $\rho \lesssim 16$, the search fails. In the intermediate range of SNRs, there is a finite detection probability less than one. 
\begin{table}
    \caption{Highest $\Upsilon_{N=1}$ from multiple searches conducted with the fiducial signal in Table \ref{tab:GW150914-like_injection}, scaled to different SNR by varying $D_L$. The values in this table should be compared against the distribution of background triggers shown in the top panel of Fig.~\ref{fig:real_false_alarm_prob_combined_figure},
    confident candidates with FAP $\lesssim 1/100$ are shaded green, these represent successful searches. The cases where a confident candidate was not found are shaded in red. The corresponding match between the search result (waveform) against the injected source is shown in Table \ref{tab:Best_matches}.}
    \label{tab:Best_upsilons}
    \vspace{2pt}
    \begin{tabular}{c|cccc}
        \hhline{|=|= = = =|}
        SNR $\rho$ & Run 1 & Run 2 & Run 3 & Run 4 \\
        \hline
        $38$ & \cellcolor{green!25}$1450.1$ & \cellcolor{green!25}$1390.3$ &  \cellcolor{green!25}$1422.3$ & \cellcolor{green!25}$1434.1$ \\
        $22$ & \cellcolor{green!25}$512.4$ & \cellcolor{green!25}$526.6$ &  \cellcolor{green!25}$605.3$ & \cellcolor{green!25}$490.5$ \\
        $20$ & \cellcolor{green!25}$404.6$ & \cellcolor{green!25}$408.3$ &  \cellcolor{green!25}$393.0$ & \cellcolor{green!25}$376.3$ \\
        $19$ & \cellcolor{red!25}$64.1$ & \cellcolor{green!25}$224.2$ &  \cellcolor{green!25}$440.5$ & \cellcolor{red!25}$67.6$ \\
        $18$ & \cellcolor{green!25}$396.7$ & \cellcolor{red!25}$34.4$ &  \cellcolor{green!25}$175.3$ & \cellcolor{green!25}$297.5$ \\
        $17$ & \cellcolor{green!25}$304.8$ & \cellcolor{green!25}$246.8$ &   \cellcolor{green!25}$225.1$ & \cellcolor{red!25}$48.5$ \\
        $16$ & \cellcolor{red!25}$56.4$ & \cellcolor{red!25}$49.0$ &  \cellcolor{green!25}$134.7$ & \cellcolor{red!25}$66.2$ \\
        
        \hhline{|=|= = = =|}
    \end{tabular}
\end{table}
\begin{table}
    \caption{Matches against injected waveform for the highest significance triggers presented in Table \ref{tab:Best_upsilons}. Matches greater than $0.97$ are shaded in green, those in the range $[0.5,0.97]$ are shaded in yellow and those below $0.5$ are highlighted in red. }
    \label{tab:Best_matches}
    \vspace{2pt}
    \begin{tabular}{c|cccc}
        \hhline{|=|= = = =|}
        SNR $\rho$ & Run 1 & Run 2 & Run 3 & Run 4 \\
        \hline
        $38$ & \cellcolor{green!25}$0.9991$ & \cellcolor{green!25}$0.9985$ &  \cellcolor{green!25}$0.9988$ & \cellcolor{green!25}$0.999$ \\
        $22$ & \cellcolor{green!25}$0.9976$ & \cellcolor{green!25}$0.9874$ &  \cellcolor{green!25}$0.9982$ & \cellcolor{green!25}$0.9987$ \\
        $20$ & \cellcolor{green!25}$0.9948$ & \cellcolor{green!25}$0.9921$ &  \cellcolor{green!25}$0.9976$ & \cellcolor{green!25}$0.9995$ \\
        $19$ & \cellcolor{red!25}$0.003$ & \cellcolor{yellow!25}$0.7789$ &  \cellcolor{green!25}$0.9993$ & \cellcolor{red!25}$0.2017$ \\
        $18$ & \cellcolor{green!25}$0.9962$ & \cellcolor{red!25}$0.0144$ &  \cellcolor{yellow!25}$0.7078$ & \cellcolor{green!25}$0.9984$ \\
        $17$ & \cellcolor{green!25}$0.9938$ & \cellcolor{green!25}$0.9949$ &   \cellcolor{yellow!25}$0.768$ & \cellcolor{red!25}$0.0$ \\
        $16$ & \cellcolor{red!25}$0.2424$ & \cellcolor{red!25}$0.1262$ &  \cellcolor{yellow!25}$0.553$ & \cellcolor{red!25}$0.1634$ \\
        
        \hhline{|=|= = = =|}
    \end{tabular}
\end{table}

\end{document}